\documentclass[final,1p,times]{elsarticle}

\usepackage{graphics}
\usepackage{graphicx}

\usepackage{amssymb}

\usepackage{multirow}
\usepackage[thmmarks]{ntheorem}
\usepackage{color}
\usepackage{amssymb,amsmath}
\usepackage{latexsym}   

\newcommand{\ie}{{\it i.e.}}
\newcommand{\eg}{{\it e.g.}}
\newcommand{\ea}{{\it et al.}}

\newcommand{\ha}{{Hadoop }}
\newcommand{\haa}{{Hadoop}}

\newcommand{\map}{\mbox{{\sf map}}}
\newcommand{\reduce}{\mbox{{\sf reduce}}}

\newcommand{\merge}{\mbox{{\sf merge}}}
\newcommand{\shuffle}{\mbox{{\sf shuffle}}}
\newcommand{\sort}{\mbox{{\sf sort}}}
\newcommand{\readmap}{\mbox{{\sf read-map}}}
\newcommand{\spill}{\mbox{{\sf spill}}}
\newcommand{\copymap}{\mbox{{\sf copy-map}}}
\newcommand{\reducewrite}{\mbox{{\sf reduce-write}}}

{
\theoremheaderfont{\normalfont\bfseries}
\theorembodyfont{\slshape}
\theoremseparator{.}
\theoremsymbol{\rule{1ex}{1ex}}

}

{
\theoremheaderfont{\normalfont\bfseries}
\theorembodyfont{\slshape}
\theoremseparator{.}
\theoremsymbol{\rule{1ex}{1ex}}

}

{
\theoremheaderfont{\sc}
\theorembodyfont{\upshape}
\theoremstyle{nonumberplain}
\theoremseparator{.}
\theoremsymbol{\textsc{(q.e.d)}}

}

{ \theoremheaderfont{\normalfont\bfseries}
\theorembodyfont{\slshape} \theoremseparator{.}
\theoremsymbol{\rule{1ex}{1ex}}

}

{
\theoremheaderfont{\sc}
\theorembodyfont{\upshape}
\theoremseparator{.}
\theoremnumbering{arabic}
\theoremsymbol{\textrm{(q.e.d)}}

}

{
\theoremheaderfont{\normalfont\bfseries}
\theorembodyfont{\slshape}
\theoremseparator{}
\theoremnumbering{arabic}
\theoremsymbol{\ensuremath{\Box}}

}

{
\theoremheaderfont{\sc}
\theorembodyfont{\slshape}
\theoremseparator{}
\theoremnumbering{arabic}
\theoremsymbol{\rule{1ex}{1ex}}

}

{
\theoremheaderfont{\normalfont\bfseries}
\theorembodyfont{\slshape}
\theoremseparator{.}
\theoremnumbering{arabic}
\theoremsymbol{\ensuremath{\Box}}

}

{
\theoremheaderfont{\normalfont\bfseries}
\theorembodyfont{\slshape}
\theoremseparator{}
\theoremnumbering{arabic}
\theoremsymbol{\ensuremath{\Box}}
\newtheorem{remark}{Remark}
}

{
\theoremheaderfont{\sc}
\theorembodyfont{\upshape}
\theoremseparator{.}
\theoremnumbering{arabic}
\theoremsymbol{\ensuremath{\Box}}
\newtheorem{example}{Example}
}

\newcommand {\argmin}[1]{\underset{#1}{\rm{argmin}}}

\journal{}

\begin{document}

\begin{frontmatter}



\title{Measuring the Optimality of \ha Optimization}

\author[PSU]{Woo-Cheol Kim}
  \ead{wxk11@psu.edu}
  \address[PSU]{College of IST, The Pennsylvania State University, University Park, PA, 16802, USA}

\author[SKK]{Changryong Baek}
  \ead{crbaek@skku.edu}
  \address[SKK]{Department of Statistics, Sungkyunkwan University, Seoul, 110-745, Korea}

\author[PSU]{Dongwon Lee\corref{cor}}
  \ead{dongwon@psu.edu}
  \cortext[cor]{Corresponding author}

\begin{abstract}
In recent years, much research has focused on how to optimize \ha jobs.
Their approaches are diverse, ranging from improving HDFS and \ha job scheduler to optimizing parameters in \ha configurations.
Despite their success in improving the performance of \ha jobs,
however, very little is known about the limit of their optimization performance.
That is, how {\em optimal} is a given \ha optimization?
When a \ha optimization method $X$ improves the performance of a job by $Y$\%,
how do we know if this improvement is as good as it can be?
To answer this question, in this paper, we first examine the ideal best case, the lower bound, of running time for \ha jobs
and develop a measure to accurately estimate how optimal a given \ha optimization is with respect to the lower bound.
Then, we demonstrate how one may exploit the proposed measure to improve the optimization of \ha jobs.
\end{abstract}


\end{frontmatter}


\section{Introduction}

With the emergence of ``big data" in many science and engineering fields, the needs rapidly arise to be able to process and analyze such big data to derive novel findings.
Addressing such needs, the Apache Hadoop has become a popular software solution, providing an efficient platform for big data analytics.
Based on Google's MapReduce~\cite{Dean:2004:MSD} paradigm, \ha delivers distributed and parallel data processing to developers in a seamless manner.

Despite its increased popularity and huge potential, using \ha is not without challenges.
One of the challenges is that it is non-trivial to {\em optimize} \ha jobs.
In general, there are many factors affecting the performance of \ha such as Hardware, \ha implementation, job algorithm, and characteristics of input/out data.
Toward these challenges, the research community has recently presented diverse solutions, ranging from improving HDFS and \ha job scheduler to optimizing parameters in \ha configurations
(\eg,~\cite{Khoussainova:2012:PDM, Dittrich:2012:EBD, Cherkasova:2011:PMM, Jiang:2010:PMI, Kwon:2012:SMS, Lee:2012:PDP, Ranger:2007:EMM}).
Although these solutions greatly succeeded in improving the performance of \ha jobs,
we believe that two important issues are largely under-addressed:
(1) Existing optimization solutions mostly focus on the optimization of \ha framework itself or algorithms running within \haa, but often ignore Hardware utilization of a \ha job.
For instance, even if a \ha job is claimed to be fully optimized, if its runtime CPU utilization is low, then there may be still some room available for improvement;
and (2) Existing solutions provide ingenious methods to improve the performance of \ha job by some data;
however, they seldom provide information about the lower bound of the improvement.
Specifically, what is the ideal performance limit that a \ha job can achieve at best?.
Such lower bound information can effectively guide developers during the optimization.

\begin{figure}
\centering
\includegraphics[width=1.0\columnwidth, clip]{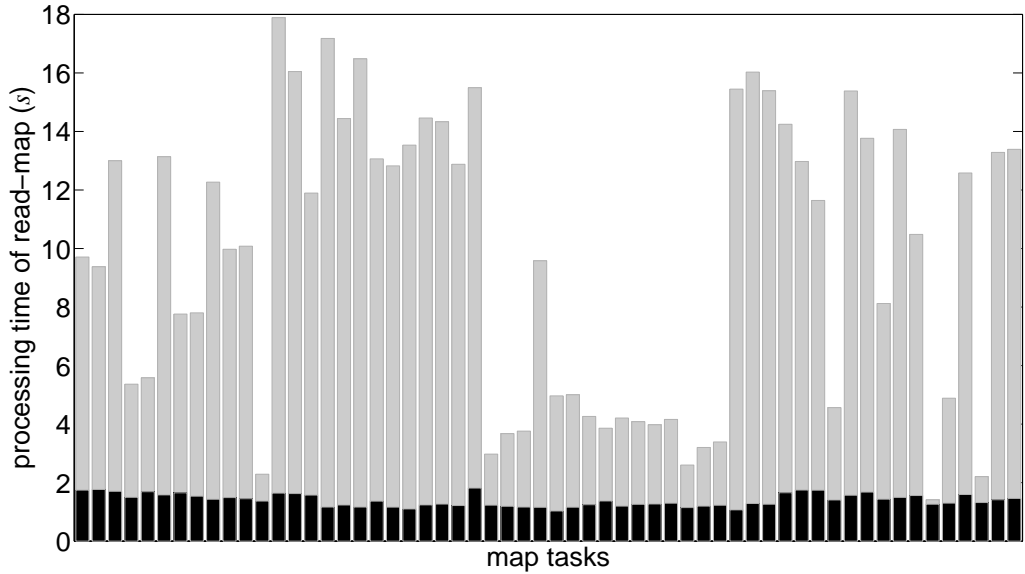}
\caption{
The actual processing time of the \readmap\ sub-phase of a \ha job after optimized by Starfish, \textit{gray bar}
, with its ideal processing time defined by this paper, \textit{black bar}.}
\label{f:comparison}
\end{figure}

\begin{example}
Figure~\ref{f:comparison} contrasts two running times:
(1) the gray bar is the running time of a \ha job after optimized by Starfish~\cite{Babu:2010:TAO}, a well-known \ha optimizer,
and (2) the black bar is the ``estimated" lower bound of the running time by our proposal.
The data were obtained by running TeraSort on a cluster with 5 nodes (\ie, 1 master node and 4 slave nodes), 2 map slots, and 2 reduce slots, and all other parameters optimized by Starfish.
Note that despite the aggressive optimization done by Starfish, potentially there is an ample room for further improvement, regardless of whether it can be done by Starfish or not.
\end{example}

To be able to estimate such a limit of \ha performance accurately, we obtain detailed profile information of each record processing in \ha jobs striking the balance between the estimation accuracy and profiling overhead, identify the cut-off threshold for outlying processing times using statistical methods, and estimate the ideal total running time by extrapolating the normal processing time of a record.

Overall, we make the following contributions.
\begin{itemize}
\item We thoroughly examine the impact of the Hardware factor toward \ha performance and investigate how to estimate the lower bound of running time accurately.

\item We present how to estimate the overheads of \ha processing time, and propose a method to estimate the ideal processing time for \map\ and \reduce\ tasks.

\item By combining estimation methods for \map\ and \reduce\ tasks, we propose a measure, $vet_{job}$, to indicate the degree of optimization of a \ha job (\eg, a job with $vet_{job}=1$ is viewed as the perfectly optimized job).

\item Using real experiments performed on diverse cluster settings and \ha configurations, we demonstrate the effectiveness of our proposed measure with respect to its accuracy, and discuss how to use our proposal to complement existing solutions such as Starfish.

\end{itemize}

Note that while we investigate ``if" there is still room for optimization for a given \ha job (and what the limit is if so),
we do {\em not} tackle ``how" to achieve such an optimization automatically.
We believe that knowing the lower bound of the optimization will be quite useful for \ha practitioners
to tune the performance of their \ha jobs (either manually or using other tools).
We plan to investigate the ``how" part (\ie, automatic means to achieve a \ha optimization close to the estimated ideal limit) in a future work.

\section{Preliminaries}

\subsection{\ha Basics}

\ha consists of distributed data storage engine (\eg, HDFS, HBase) and MapReduce execution engine,
and has been successfully used for processing highly distributable problems
across a large amount of datasets using a large number of nodes.
These nodes, called a \ha cluster, in turn consist of a single master node and multiple worker nodes.
In this framework, a user program is called a {\em job}, and is divided into two steps, the {\bf map} and {\bf reduce}.
\begin{itemize}
\item Map step: The master node (\ie, JobTracker in \haa) takes the input data, partitions it into smaller sub-problems, called \map\ tasks, and distributes them to the worker nodes (\ie, TaskTracker in \haa).
Each worker node then processes the \map\ tasks, and passes signals back to its master node when the processing is done.
\item Reduce step:
The master node allocates all results from the \map\ step to worker nodes. Then,
the \reduce\ task in each worker node collects the answers to all \map\ tasks and combines them in some way to derive the answer to the problem.
\end{itemize}


\begin{figure}[tb]
\centering
\includegraphics[width=1.0\columnwidth, trim=5mm 185mm 45mm 12mm, clip]{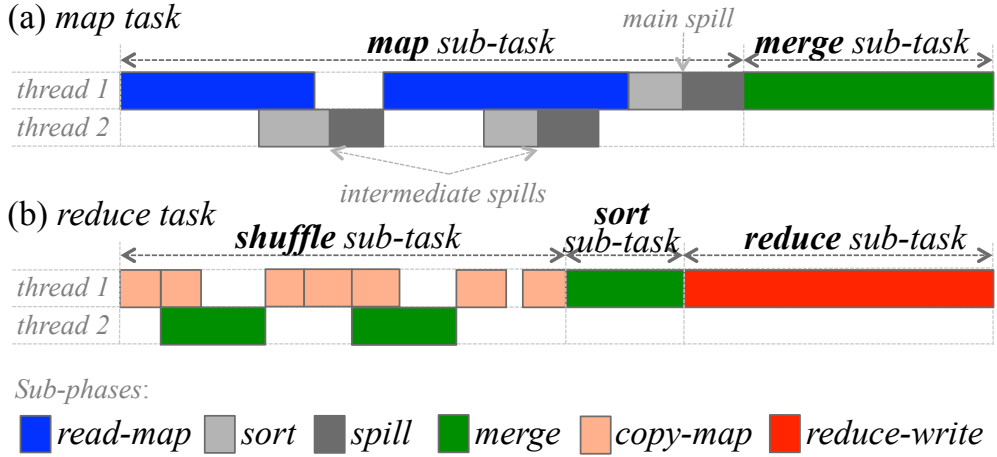}
\caption{The illustration of task, sub-task, and sub-phase in a \ha job:
(a) a \map\ task consists of \map\ and \merge\ sub-tasks; and
(b) a \reduce\ task consists of \shuffle, \sort, and \reduce\ sub-tasks.}
\label{f:subphases}
\end{figure}

In analyzing \ha jobs, people often use terms with different granularities--\ie, {\bf task}, {\bf sub-task}, and {\bf sub-phase}.
While the \map\ task consists of 2 sub-tasks (\ie, \map\ and \merge), logically,
the \reduce\ task consists of 3 sub-tasks (\ie, \shuffle, \sort, and \reduce).
Each sub-task in turn gets divided into many sub-phases such as \readmap, \spill, \merge, \copymap, and \reducewrite.
Figure~\ref{f:subphases} illustrates these terms and their relationships.

\subsection{\ha Parallel Processing}


\ha provides 3-levels of parallelism:
(1) multi-nodes within a \ha cluster,
(2) multi-slots within a node, and
(3) multi-threading within a task.
In general, each node has multiple slots for both \map\ and \reduce\ tasks.
The number of slots typically indicates how many tasks can run concurrently.
Often sub-tasks read a file from and to a local disk or network disk through the HDFS protocol.
Intermediate output of all sub-tasks, especially the \map\ and \shuffle\ sub-tasks, is stored in a memory buffer to reduce disk I/O.
However, since the size of the output may exceed that of memory buffer, when such an overflow occurs, \spill\ is needed to flush the data into a local file system.
This \spill\ is often implemented using multi-threading to maximize the utility of disk I/O and to reduce the running time of jobs.

Often, the ``rule of thumbs" is to set the number of slots per node as the number of processor cores in a node to maximize parallel processing.
If there are tasks that run in multi-threads at some point, some of threads will share a CPU core and disk I/O and become slower (unless they were scheduled on different cores), due to the context-switching overhead.
Multi-threading also causes more disk I/Os which must be processed for a short time.
This will create a bottleneck of job execution if a disk in a node does not support the I/O requests fully.
The performance of these parallel processing is directly related to available hardware resources.
If the cluster does not support enough resources for multi-threads,
for instance, the running time of some sub-phases running in multi-threads would be longer than it would be under single-thread.
In real settings, since such cases are not uncommon, being able to understand dynamic interplay and relationships among multi-threads and multi-tasks of \ha jobs is important.

While the 3-level parallelism is one of the key reasons for the success of \haa, it is also the culprit for the difficulty of \ha optimization.

\subsection{\ha Cost Model}

Although the running time of a job is affected by many factors,
Starfish~\cite{Babu:2010:TAO,Shivanth:2011:PWC} shows that the following are major factors:
(1) Hardware (or resources) such as CPU Clock, disk I/O, network bandwidths, and memory size;
(2) Job algorithms (or program) such as map, reduce, partition, combine, compress;
(3) Data size for input data, shuffle data, and output data, which are closely correlated with the running time of a job; and
(4) \ha configuration (or parameters).
This last factor usually determines the running time of a job since the other three factors are {\em not} modifiable after the \ha cluster is set up and the job starts the execution.
Based on the four factors, for instance, one can discover the best parameter configuration or cluster set-up as an optimization problem (as Starfish successfully did).
The simple default profiler in \ha provides basic information such as start/end time, input/output data size, and number of map/reduce tasks.

The running time of both \map\ and \reduce\ tasks is highly correlated with that of \textit{job}.
This is because parameters such as the number of nodes in a cluster or the number of slots in a node are un-modifiable in a typical environment.
Therefore, optimizing \ha jobs is often equivalent to optimizing both \map\ and \reduce\ tasks.
In general, however, the optimization of \ha jobs is a hard problem,
as a \ha job is often composed of many sub-modules, implementing different algorithms.
Some of these sub-modules are connected in serial, while others are connected in parallel.

\section{The Best Scenario} \label{s:hadoop}

In this section, we discuss what the ``best scenario" means in running a \ha job.

\subsection{The Platform Best Scenario} \label{sec-platform}

When the multiple intermediate \spill s occur, they are first saved in a local storage due to insufficient memory buffer during a \map\ sub-task, and later merged into a single file to save time for transfering an output of a \map\ task into a \reduce\ task.
For example, in Figure~\ref{f:subphases}(a), there occur two intermediate \spill s and a main \spill\ during a \map\ task, which are then combined into a single file in the subsequent \merge\ sub-task.
Note that in general, if there were {\em no} intermediate \spill\ sub-phases, then the subsequent \merge\ sub-task would not be needed (since there is nothing to merge).
In such a case, a \map\ sub-task runs on a single thread.

As a result, this no-intermediate spill scenario yields two advantages in \map\ task optimization:
(1) It eliminates unnecessary \merge\ sub-tasks; and
(2) It increases the utilization of the main thread for processing \readmap\ sub-task, blocking other threads for spilling the intermediate output in a memory into a disk.
In this paper, we refer to this scenario of having one main \spill\ without any intermediate \spill\ as the {\em platform best scenario}, as also noted in~\cite{Todd:2012:OMJ}.
The number of spills depends on the size of memory buffer during the \map\ task.
Multiple parameters of Hadoop are associated with the size of memory buffer (\eg, \textit{io.sort.mb}, \textit{io.sort.spill.percent}, \textit{mapred.child.java.opts}, and \textit{io.sort.record.percent}).
Starfish also uses these parameters to optimize Hadoop jobs.
According to our preliminary experiments, Starfish optimization often results in a single main \spill\ in a \map\ task without any intermediate \spill s, coinciding with our definition of the platform best scenario.

If the size of output data from a \map\ task is bigger than that of memory buffer set by the user, it is impossible to achieve the platform best scenario.
However, such a case is rather rare, considering the modern server specification.
For example, Hardware specification of ``M1 Medium Instance'' of AWS includes 2 CPU cores and 3.75 GB memory.
Considering the fact that the number of \map\ tasks and \reduce\ tasks running at the same time is the same as the number of cores in a node, the available memory size for a \map\ task is over 1 GB at least even if OS consumes 1.75 GB memory.
Note that the typical size of input data used for a \map\ task is 128 MB and the size of output data is usually no greater than that.
The platform best scenario can be, for instance, easily fulfilled when we allocate 200 MB out of 1 GB as the memory buffer.

The \sort\ sub-task in a \reduce\ task is similar to the \merge\ sub-task in a \map\ task.
In addition, the \merge\ sub-phase within a \shuffle\ sub-task in a \reduce\ task is similar to the intermediate \spill\ sub-phase within a \map\ sub-task in a \map\ task.
In general, the number of intermediate \spill s in a \map\ task can be easily controlled by changing the Hadoop parameters.
However, it is difficult to control the number of \merge\ sub-phases in a \reduce\ task
because it is affected not only by the size of the memory buffer but also by the number of \map\ tasks that send the output data to the \reduce\ task.
It is on un-modifiable parameter and usually depends on the total size of input data in a \textit{job}.
Therefore, in this paper, we focus on optimization of the \map\ task.

\begin{remark}
When a \map\ task incurs a single main \spill\ sub-phase and no intermediate \spill s, the running time of the \map\ task becomes the minimum.
\end{remark}

\begin{remark}
When a \reduce\ task incurs a minimum number of the \merge\ sub-phase, the running time of the \reduce\ task becomes the minimum.
\end{remark}

\subsection{The Empirical Best Scenario}

In general settings, the sizes of input and output data of \map\ tasks in a job are usually the same.
Although it depends on the specific type and implementation of the job, this assumption usually holds for Hadoop job analysis.
Under this assumption, if one finds the Hadoop parameters satisfying the platform best scenario,
the running times of all \map\ tasks are expected to be very similar.
However, in real settings, often this is not the case.
A significant skew among the running times of \map\ tasks has been reported~\cite{Kwon:2012:SMS}.
In general, the slowest \map\ task takes orders of magnitude longer to finish than the fastest \map\ task.
The major reason for such a skew is that both \map\ and \reduce\ tasks share insufficient hardware resources.

Consider a normal Hadoop cluster setting.
Based on the rule of thumb, we set the number of slots per node as the number of processor cores in a node to maximize parallel processing.
If there are tasks that are running on multi-threads at some point,
some threads will share a CPU core and disk I/O, running slower than
if they were scheduled on different cores, due to thread context switching overhead.
Most of jobs are known as I/O bounded.
Multi-threading also causes more disk I/Os, creating a bottleneck of job execution
if a disk in a node does not fully support the I/O requests.
The performance of parallel processing is directly related to the hardware resources available to use.
If the cluster does not support enough resources for multi-threads, the running time of some sub-phases running on multi-threads will be longer than when they run on single-thread.
In real settings, such cases occur frequently.


The most important hardware resources associated with the processing time of a task are CPU and I/O devices such as disk and network.
If each \map\ task is running on its own CPU and I/O devices supporting it fully,
the processing time of all \map\ tasks will be the same and the minimum.
We refer to such a scenario as the {\em empirical best scenario} and the minimum running time as the {\em ideal \map\ time.}
In real settings, however, a \map\ task cannot have a CPU core exclusively, due to the multi-threading implementation of \haa.
As a result, context switching overhead for multi-threading should be added to the ideal map time.
This is known as the \textit{CPU overhead}.
The second overhead type, \textit{I/O overhead}, comes from disk and network.
A \map\ task reads input data and write output data from/to disk or network when HDFS is used.
These I/O devices sometimes block CPU when I/O requests of a \map\ task need to wait until previous requests are processed.

\begin{remark}
The real processing time of a \map\ task is defined as the sum of the ``ideal \map\ time $+$ CPU overhead $+$ I/O overhead."
\end{remark}

\section{Measuring the Optimization} \label{s:opt-measure}

In this section, we discuss how the overheads can be measured and present our method to measure the optimization of \haa.



\subsection{Estimating Ideal \map\ Time} \label{s:map-time}

The processing time to finish a \map\ task consists of several cost components.
Among such components, there are costs starting a task in the beginning and cleaning a task at the end.
These costs are largely independent of job type.
For instance, the cost to clean up a task is mostly waiting time for messages from/to the master node--\eg, sending a ``ready" message.
In general, since it is hard to predict or control such costs,
existing approaches such as \cite{Shivanth:2011:PWC} exclude these costs when comparing ideal \map\ time.
We adopt the same approach and ignore the costs for starting/cleaning sub-tasks.

Besides the time for starting/cleaning sub-tasks, then, time to process a \map\ task mainly consists of time for a \map\ sub-task and a \merge\ sub-task.
However, as discussed in the platform best scenario, the \merge\ sub-task will not occur if all \map\ tasks run with only ``one" \spill.
Therefore, the time for \map\ sub-task is the dominating factor for the overall time for \map\ task.
In turn, the time for \map\ sub-task can be split to two components: time for \spill\ sub-phase and that for \readmap\ sub-phase.

\begin{figure}
\includegraphics[width=0.49\columnwidth]{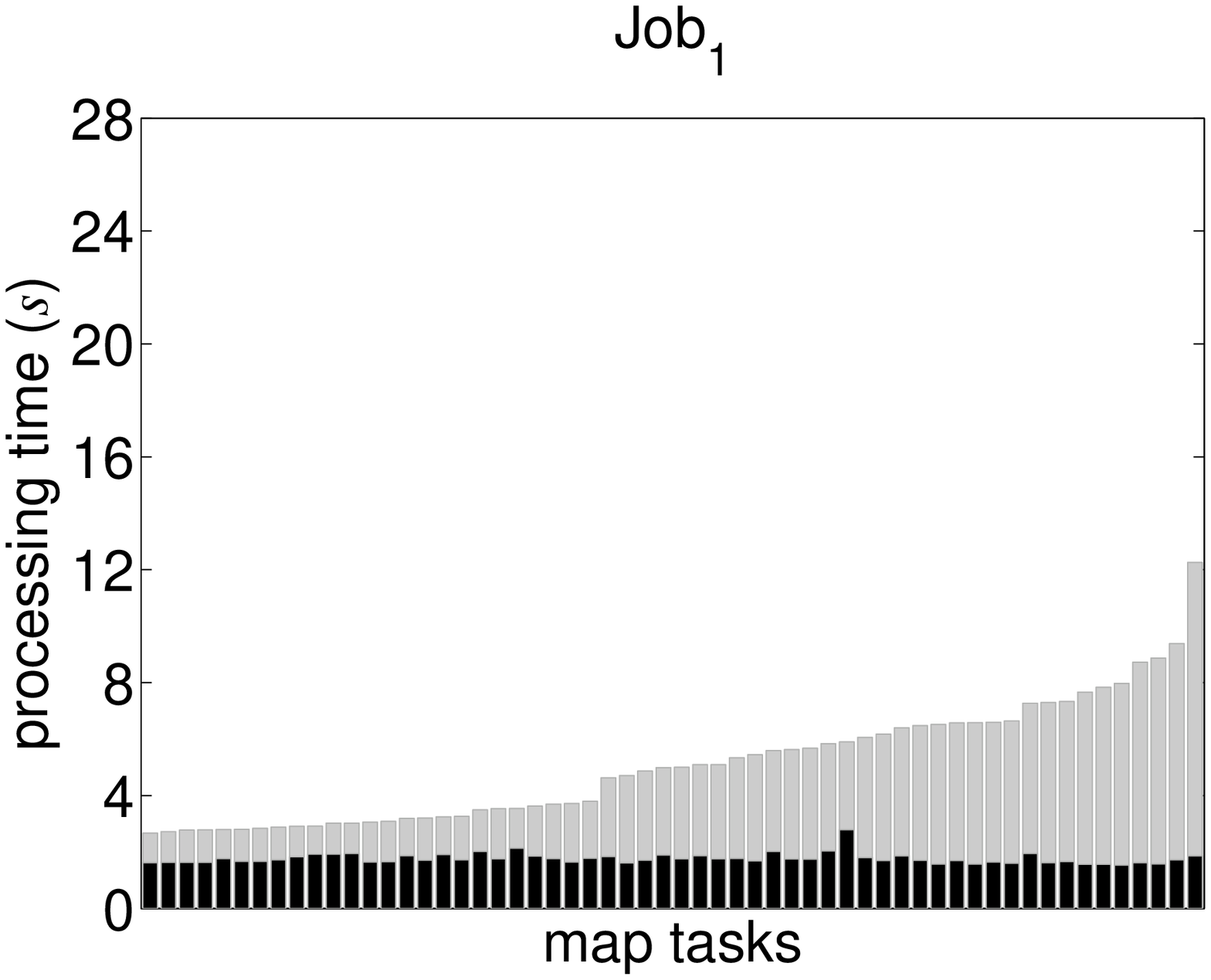}
\includegraphics[width=0.49\columnwidth]{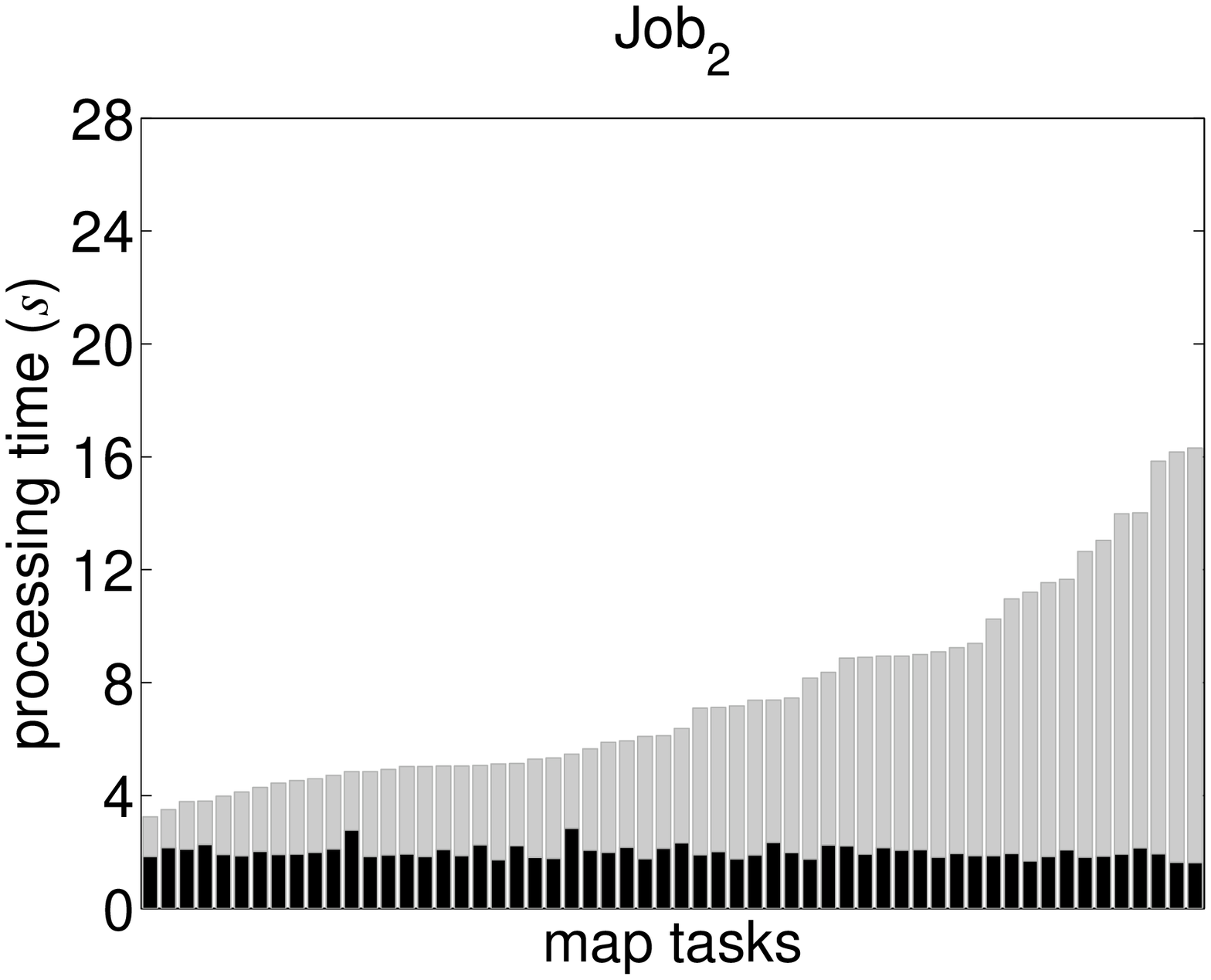}
\includegraphics[width=0.49\columnwidth]{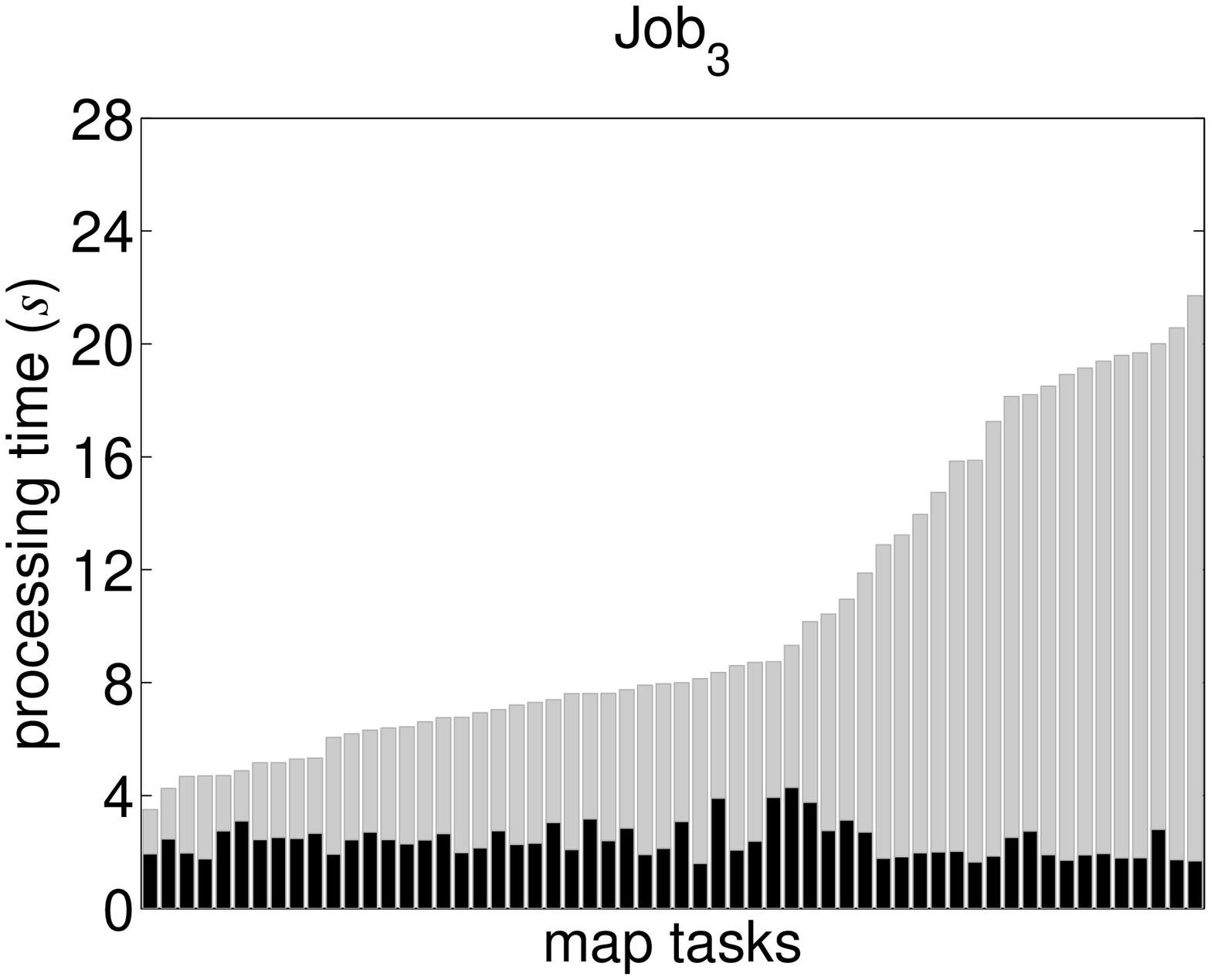}
\includegraphics[width=0.49\columnwidth]{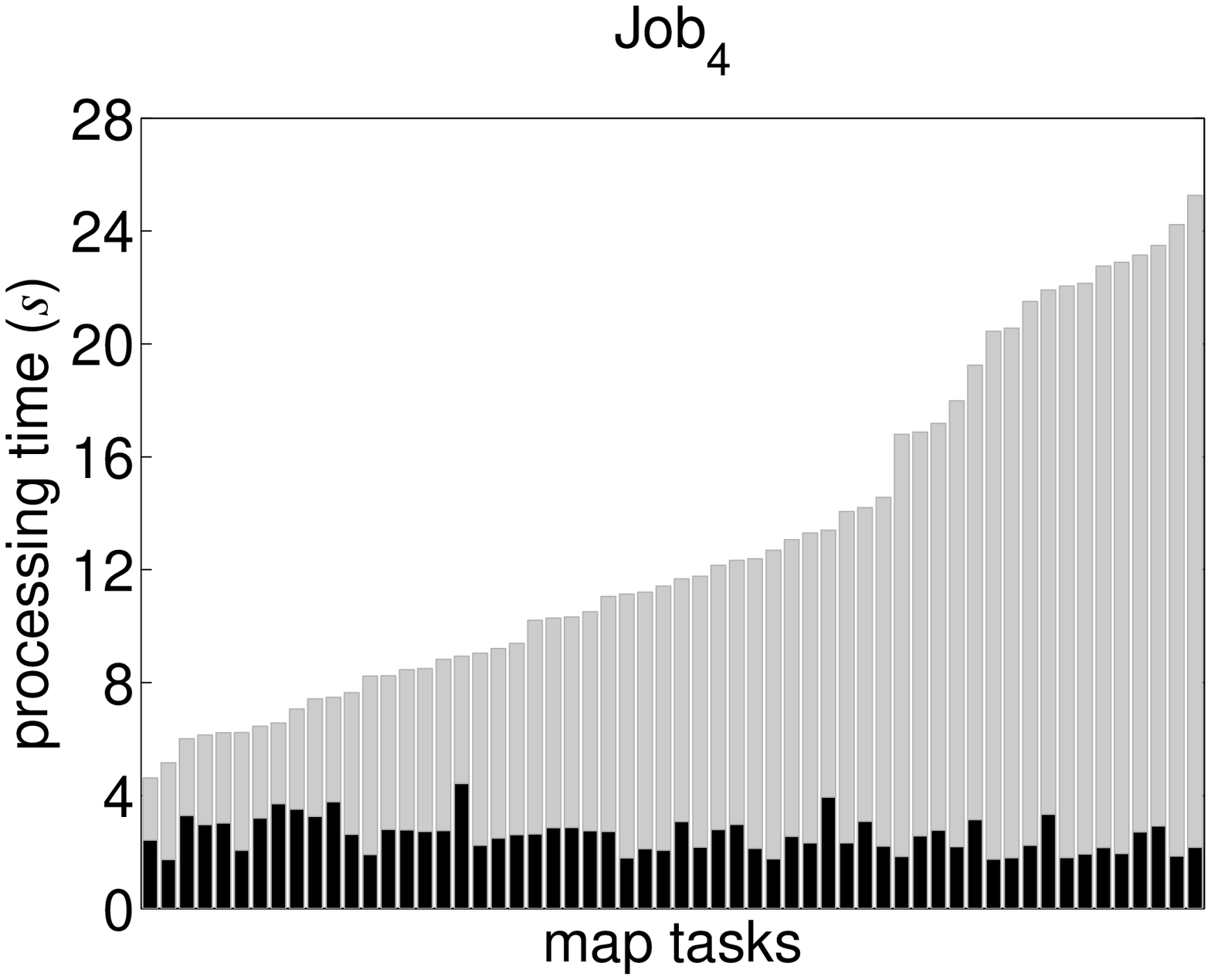}
\caption{
The processing times of \spill\ sub-phase using the configuration of jobs in Table~\ref{t: job confs}.
The \textit{gray bar} and \textit{black bar} are the processing times of \map\ task and \spill\ sub-phase within the corresponding \map\ task, respectively.
The $X$-axis plots the IDs of \map\ tasks.
Note that for all four jobs, the processing times of \spill\ sub-phases remain relatively constant,
regardless of \map\ tasks to which each belongs.}
\label{spill_dura}
\end{figure}

In general, the processing time of the \spill\ sub-phase consists of time to sort the collected records by record keys and time to flush them out from a memory to a storage.
While it is {\em not} straightforward to measure the processing time of a record in a \spill\ sub-phase,
its characteristics are such that:
(1) the processing time of the \spill\ sub-phase is much smaller than that of the \readmap\ sub-phase and
(2) its processing times do not change much across different \map\ tasks.
For instance, as shown in Figure~\ref{spill_dura},
for all four jobs, the processing times of \spill\ sub-phases remain relatively constant, regardless of \map\ tasks to which each belongs.
Therefore, we further decided to exclude the time for the \spill\ sub-phase in estimating the time for the \map\ sub-task.
In order words, the processing time for \map\ task is dominated by the time for \map\ sub-task,
that in turn is dominated by the time for \readmap\ sub-phase.


\subsection{Estimating Ideal \readmap\ Time} \label{s:readmap}

One of the important features of \ha is that all data are processed in a unit of {\em record}.
In theory, since each record has almost the same size and gets processed by the same user-defined algorithm,
the time to process such a record should be similar.
However, in practice, the processing times of records within a task vary significantly.
In general, the \textit{base cost} to process records in the \readmap\ sub-phase consists of three costs to:
(1) reading a record into memory,
(2) processing a record using CPU, and
(3) writing a record to memory.
Subsequently, two additional costs are added to the base cost:
(1) the unavoidable \textit{I/O cost} from disk access and spill, and
(2) the unpredictable but reducible \textit{overhead} such as I/O waiting cost resulted from heavy I/O requests or context switching in multi-threading.


\begin{figure}
\centering
\includegraphics[width=0.95\columnwidth, trim=5mm 220mm 85mm 12mm, clip]{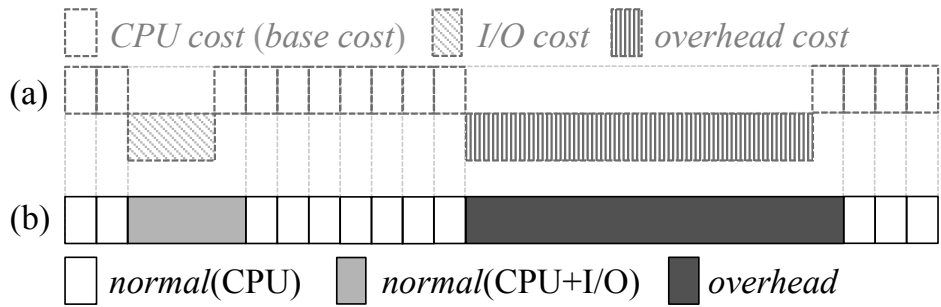}
\caption{Example of collected record processing time: (a) logically separated costs and (b) actual profiled costs}
\label{records}
\end{figure}

Figure~\ref{records} illustrates an example of record processing timeline based on our analysis,
where the base cost (\ie, CPU cost), the first additional cost (\ie, I/O cost), and the second additional cost (\ie, overhead cost) are visualized, respectively.
In theory, we can categorize a type of record processing cost as shown.
However, in practice, what we observe is only the duration between processing records.
That is, overall record processing times appear as shown in not Figure~\ref{records}(a) but Figure~\ref{records}(b).
The processing time of a record has a range from a few nanoseconds ($10^{-9}$) to a few thousand nanoseconds.
It is much smaller than the time interval between disk accesses, which is about a few milliseconds ($10^{-3}$),
and the time interval between context switching, a \textit{time quantum}, which is about a few hundred milliseconds.
As a result, only a small number of records take a much longer time to process due to unnecessary overheads, context switching, and disk access.
Most of record processing times come from memory access/read time and CPU clocks, determined by a user-defined algorithm.
Consider the following example.

\begin{figure}[tb]
\centering
\includegraphics[width=1.0\columnwidth, trim=5mm 15mm 5mm 15mm, clip]{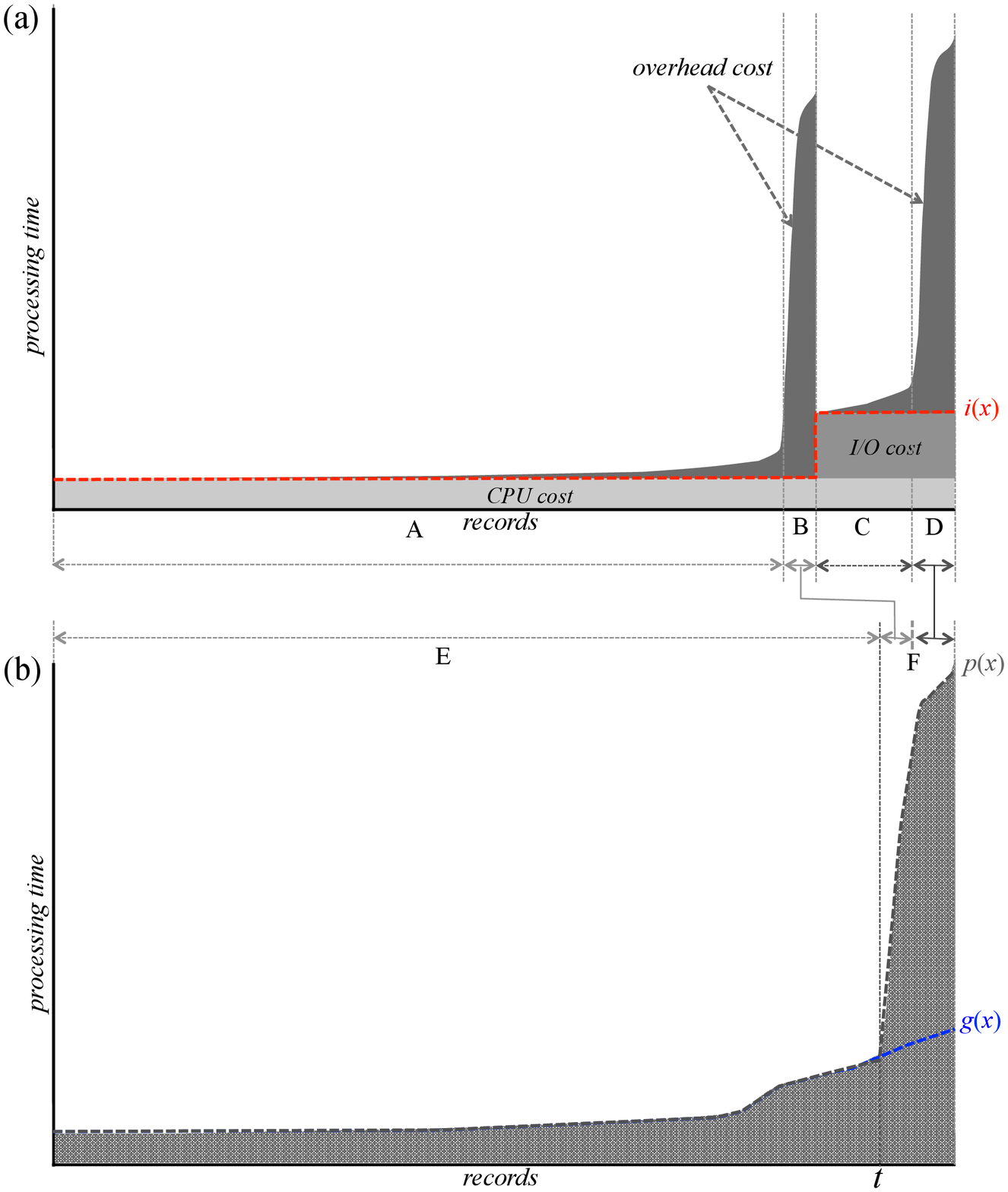}
\caption{
Distributions of record processing times for (a), a theoretical case, and (b), an actual observation.
Both {\em A} and {\em C} refer to records whose processing times involved normal (CPU) and normal (CPU+I/O) times of Figure~\ref{records}(b), respectively,
while both {\em B} and {\em D} refer to records whose processing times include some overhead from CPU or I/O.
The function, $i(x)$, red dot in (a), represents the {\em ideal} distribution of record processing times,
while the function, $p(x)$, black dot in (b), represents the {\em actual} distribution of record processing times.
Finally, the function, $g(x)$, blue dot in (b), represents the distribution of the estimated optimal record processing times.}
\label{tasks}
\end{figure}

\begin{example}
Figure~\ref{tasks}(a) shows the distribution of processing times of all records in the \readmap\ sub-phase ordered by the processing time.
We can classify the record processing cost in a \map\ task into four types, {\em A} -- {\em D}, as shown in Figure~\ref{tasks}(a):
{\em A} for the normal (CPU) in Figure~\ref{records}(b),
{\em C} for the normal (CPU+I/O) in Figure~\ref{records}(b),
while both {\em B} and {\em D} for overheads in Figure~\ref{records}(b).
Note that {\em A+B} are times using only CPU as the base cost
while {\em C+D} are ones using both CPU and I/O as the base cost.
While both {\em A} and {\em C} also include some overhead,
they are different from overheads in {\em B} and {\em D} in that the former is irreducible while the latter is reducible.
On the other hand, Figure~\ref{tasks}(b) illustrates the actual distribution of record processing times that one can gather from the profiled data.
One can discover from this graph that the records on the left side of the graph, denoted by {\em E},
have far less overhead than those on the right side, denoted by {\em F}.
That is, {\em E} is in essence the sum of {\em A} and {\em C}, having affected by overheads very little,
while {\em F} is the sum of {\em B} and {\em D}, where substantial overheads occurred.
\end{example}

Several functions are plotted in Figure~\ref{tasks}.
The function, $i(x)$, red dot in Figure~\ref{tasks}(a), is the trace line of the {\em ideal} distribution of record processing times,
excluding all overhead costs, representing the platform best scenario in Section~\ref{sec-platform}.
On the other hand, the function, $p(x)$, black dot in Figure~\ref{tasks}(b), represents the {\em actual} distribution of record processing times as one observes.
All overhead costs occurring during the record processing are piled up toward the right, yielding a peak.
Such a peak is the delay that a \ha user experiences in processing a \map\ task.
Finally, the function, $g(x)$, blue dot in Figure~\ref{tasks}(b), represents the estimated distribution of ideal record processing times,
simulating the function $i(x)$ in Figure~\ref{tasks}(a) as close as possible.
That is, in order to measure the ideal cost for a \readmap\ sub-phase, since we cannot obtain the function $i(x)$ directly, we need to identify the estimating function $g(x)$ instead.
Note that both functions $p(x)$ and $g(x)$ behave almost identically from the beginning until the time $t$ in Figure~\ref{tasks}(b) (X-axis).
Therefore, for $n$ records to process, the function $g(x)$ is defined as follows:
\[
{g}(x) =
  \begin{cases}
    p(x) & \text{if } 1 \leq x \leq t \\
    \hat{g}(x) & \text{if } t+1 \leq x \leq n
  \end{cases}
\]
where $\hat{g}(x)$ is a suitable estimate of $i(x)$, to be elaborated in the next subsection.

In this paper, we refer to the {\em area under} $p(x)$ and $g(x)$ in Figure~\ref{tasks}(b) as the {\bf PR} (profiled real) and the {\bf EI} (estimated ideal) costs of record processing time in the \readmap\ sub-phase, respectively.
In addition, we refer to the estimated {\em overhead cost} in the \readmap\ sub-phase,
\ie, the difference between the area under $p(x)$ and that under $g(x)$ in Figure~\ref{tasks}(b), as the {\bf OC}.

\subsection{Estimating $t$ and $g(x)$ in \readmap\ Sub-phase} \label{s:stat-analysis}

The first task to estimate the function $g(x)$, as defined above, is to find the time point $t$ when I/O cost starts to change dramatically.
In the statistical viewpoint, such point is called the {\em change-point} and has been well studied for decades.
One of the most popular estimators to find the change-point is by minimizing the sum of squared errors. The least squares estimation (LSE) method is known to be robust with respect to underlying distributions,
and enjoys good theoretical properties such as consistency.

Let $\{X_i\}$, $i=1, \ldots, n$, be the record processing time and denote its $i$-th smallest value, {\it i.e.} order statistic, as $Y_i = X_{(i)}$. Then, the estimated change-point, $\hat{t}$, is defined as:
\[
  \hat{t} =
    \argmin{\omega\leq k \leq n-\omega} \left\{ \sum_{i=1}^k \left( Y_i - \beta_0 - \beta_1 i \right)^2
  + \sum_{i=k+1}^n \left( Y_i - \beta_2 - \beta_3 i \right)^2 \right\}
  \label{d:change_point}
\]
where $\omega$ is a pre-specified (\eg, $\omega=3$) probing window to control the fitting for observations.
The idea is very simple.
For the fixed window, $k \in \{\omega, \omega+1, \ldots, n-\omega \}$, we prepare two sub-samples, namely $\{Y_1, \ldots, Y_k \}$ and $\{Y_{k+1}, \ldots, Y_n \}$.
Then, a linear regression line is fitted for each sub-sample, and the sum of residuals are added.
The change-point is then the index $k$ achieving the smallest sum of squared errors.
This LSE estimator can be naturally extended to a higher order local polynomial regression, \eg, $\beta_0 + \beta_1 i + \ldots + \beta_p i^p$, $ p \ge 2$, but our analysis shows that a simple linear regression seems to be sufficient.

Next, to estimate the function $g(x)$, there are two mathematical restrictions to consider:
(1) $g(x)$ should be continuous at the change-point $\hat{t}$, and
(2) $g(x)$ should be monotonically increasing since the observations are ordered.
Therefore, we consider the following simple moving average form of predictor:
\[
  \hat{g} \left( r+1 \right) = 2 \hat{g} \left( r \right) - \hat{g} \left(r-1\right), \quad r=\hat{t} \leq r \leq n
\]
with initial values $\hat{g} \left( t-1 \right) = Y_{t-1}$, $\hat{g}(t) = Y_{t}$.
The idea of such filter is based on the three point moving average in time series analysis by imposing that $ \hat{g}(r) = \frac{\hat{g}(r+1) + \hat{g}(r-1)}{2}$.
In other words, in three point moving average, middle point $\hat{g}(r)$ should always be the average and the observations are ordered, hence the latter point $\hat{g}(r+1)$ should always be monotonically increasing.

Finally, using the estimated $\hat{t}$ and $\hat{g}(x)$ function,
the estimated ideal record processing time (EI), and
the estimated overhead cost (OC) can be defined as follows:
\[
  EI = \sum_{r=1}^{\hat t} Y_r + \sum_{r= \hat t +1}^n \hat{g}(r),  \hspace{0.1in}
  OC = \sum_{r=\hat t + 1}^n \left( Y_r - \hat{g} \left( r \right) \right)
\]

\begin{figure}
\centering
\begin{tabular}{cc}
  \hspace{-0.2in}
    \includegraphics[width=0.49\columnwidth]{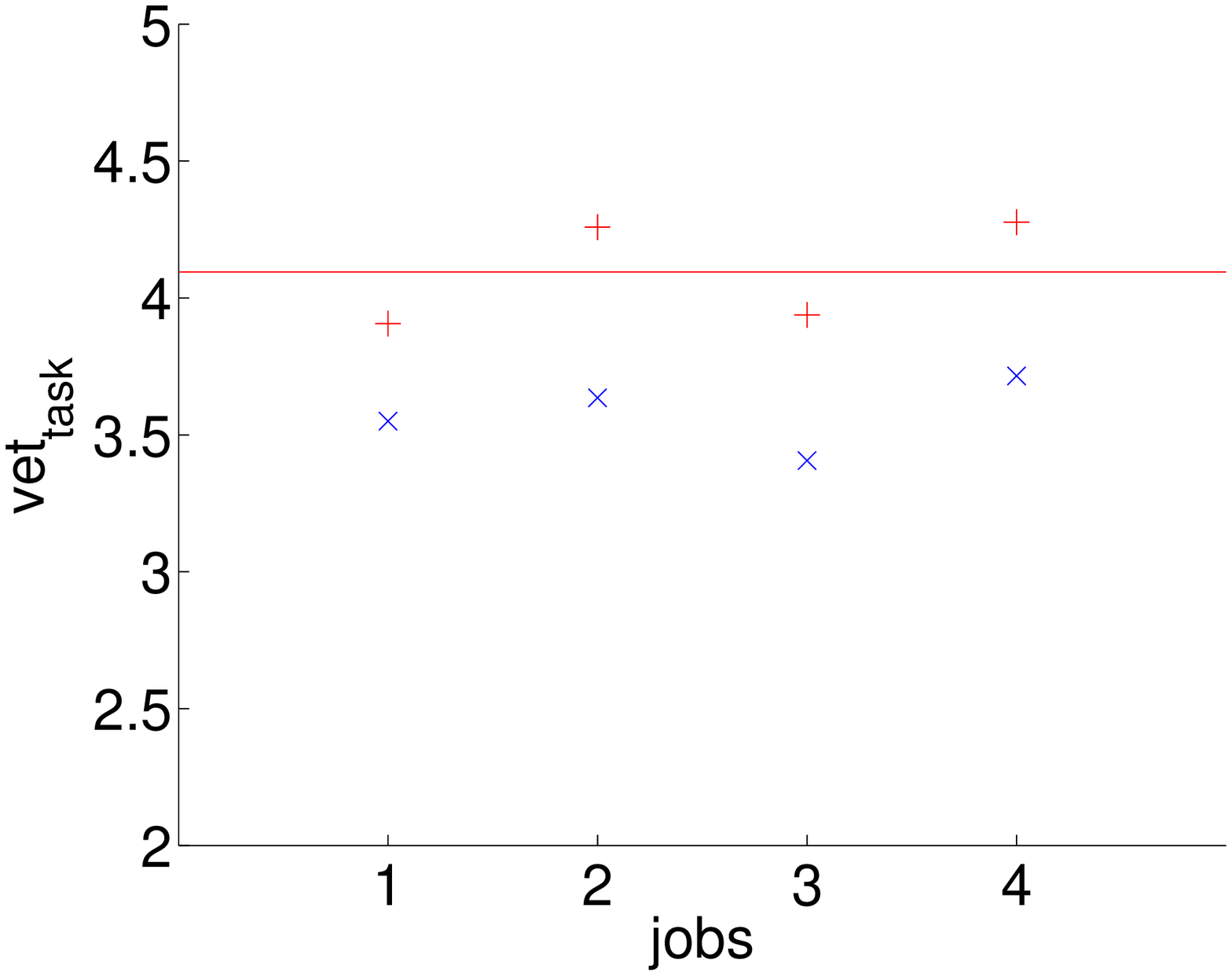} &
  \hspace{-0.2in}
    \includegraphics[width=0.49\columnwidth]{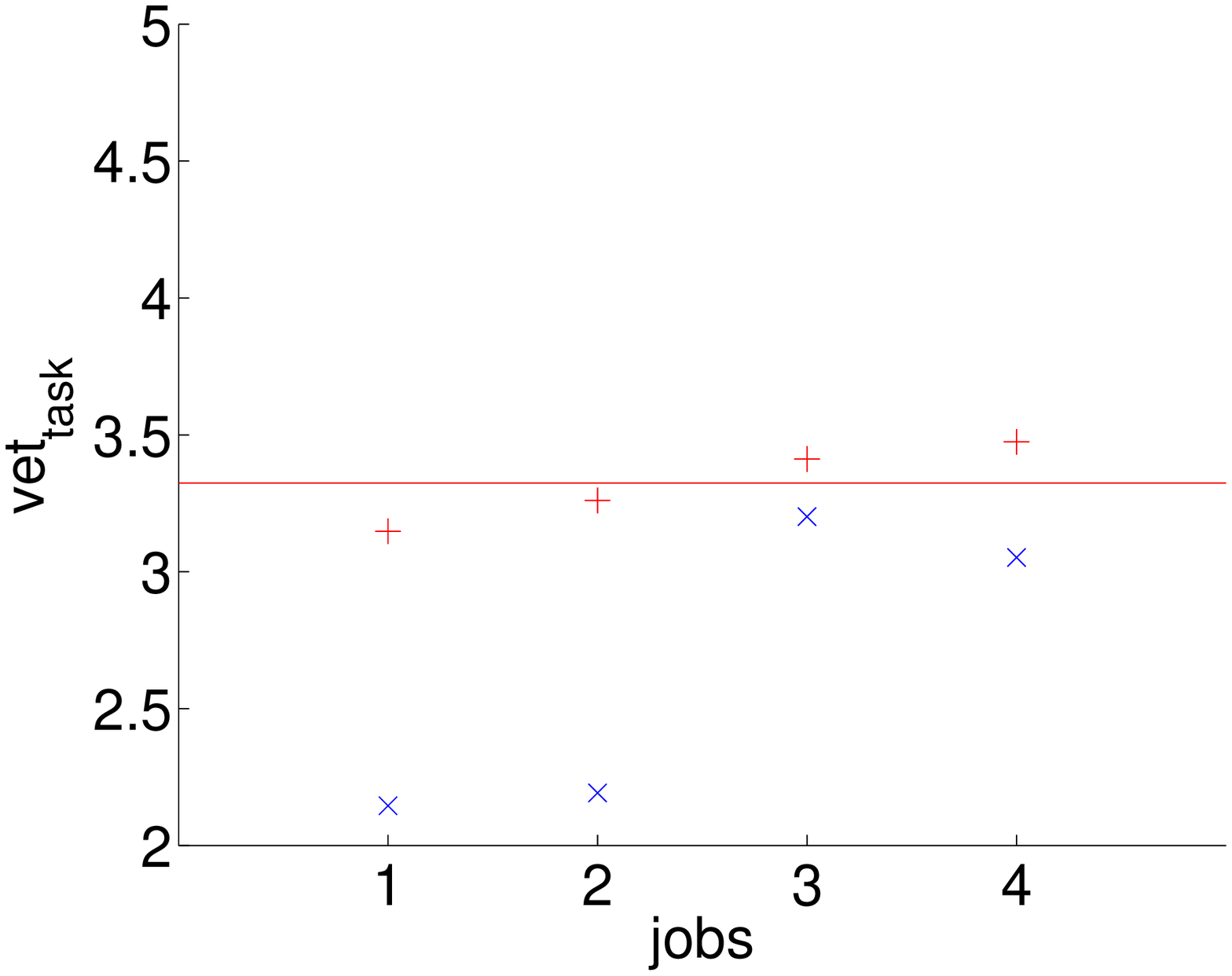} \\
    (a) With dataset A & (b) With dataset B
\end{tabular}
\caption{
The statistics of $vet_{task}$ scores with same configurations:
The red ``+'' and blue ``x'' are the sample average and the median of $vet_{task}$ of each job, respectively.
The horizontal red line is the sample average of $vet_{task}$ of four jobs in each graph.}
\label{f:ks}
\end{figure}

\subsection{Measuring the Optimization of a Job}

Now, based on all discussion so far, we are ready to define a notion to measure the degree of optimization in \haa.
Using the raw information from \readmap\ sub-phase in Sections~\ref{s:readmap} and~\ref{s:stat-analysis},
we first define {\bf vet}$_{task}$\footnote
{The name $vet$ comes from \textit{veterinarian}.
Since most names of \ha-related optimization s/w are animals
and the purpose of our measure to diagnosis them, we named it as such.}
as the measure to estimate the degree of optimization of a \ha task as follows:
\[
  vet_{task} = \frac{EI + OC}{EI}
\]
The score of $vet_{task}$ is linearly proportional to the degree of overhead in the computation
while it becomes 1 in the ideal scenario with no overhead.
Further, as a representative measure for many tasks within a job, we simply use {\bf vet}$_{job}$ as follows:
\[
  vet_{job} = \frac{1}{m} \sum_{i=1}^m {vet}_{task}^{(i)}
\]
where ${vet}_{task}^{(i)}$ is the $vet$ score for $i$-th task in a job.

Note that as shown in Figure~\ref{f:comparison}, the distribution of task processing time is skewed,
implying that larger processing times can be observed with a non-negligible probability.
Due to the bottleneck of resources in processing records in \haa,
such skewness has been reported in~\cite{Kwon:2012:SMS} as well.
Despite such skewness, however, we use the simple average of $vet_{task}$ as the representative $vet$ score.
A sample average is the simplest and most popular estimator for the center of population with good theoretical properties.
For example, the law of large numbers guarantees that sample average converges to true underlying population mean if observations are from the same population.
In other words, such skewness may produce larger processing times at record level,
but the sample average at task level is likely to be very close to true mean for moderate sample size.

To corroborate our decision to use the average, in Figure~\ref{f:ks} we compared the statistics of $vet_{task}$ of jobs running in the same environment.
Although there are some differences in the distribution of $vet_{task}$ scores of four jobs,
the differences in the ``averages" of $vet_{task}$ scores are small, especially in Figure~\ref{f:ks}(b).
In addition, we applied a formal non-parametric {\em Kolmogorov-Smirnov} (KS) statistical testing procedure
to determine whether or not two samples are from the same underlying population~\cite{massey1951kolmogorov}.
For instance, the KS test for jobs 1 and 2 in Figure~\ref{f:ks}(a) respectively yielded the $p$-value of 0.6068.
Therefore, there is no strong evidence ``against" that both samples are from the same population.

\section{Evaluations} \label{s: evaluations}

\subsection{Configuration}

We used three cluster configurations to evaluate our proposed idea as follows:
(1) \textbf{C1}: The first cluster consists of five nodes (one master node and four slave nodes), each with one quad-core CPU, 4GB of RAM, and 1TB 7200RPM of HDD. The five nodes are connected by Gigabit Ethernet in the same rack;
(2) \textbf{C2H}: The second cluster consists of a single node with two quad-core CPUs, 32GB of RAM, and 1TB 7200RPM of HDD; and
(3) \textbf{C2S}: The third cluster has the same configuration as C2H, except that it uses SSD, instead of HDD. All nodes used Ubuntu 11.4, {\haa} framework 0.20.203, and Java 1.6.0.

Among many parameters in {\haa}, we used the following configurations:
HDFS block size (\textit{dfs.blocksize}) as 64MB,
memory size for Java VM (\textit{mapred.child.java.opts}) as 200MB,
number of streams to merge at once while sorting files (\textit{io.sort.factor}) as 3,
number of map slots per node (\textit{mapred.tasktracker.map.tasks.maximum}) as 2 for C1 and 6 for C2H/C2S, and
number of reduce slots per node (\textit{mapred.tasktracker.reduce.tasks.maximum}) as 2 for all three clusters.
Other parameters concerning the number of spill (e.g., \textit{io.sort.mb}, \textit{io.sort.record.percent}, and \textit{io.sort.spill.percent}) were set according to input data size and job type,
making the number of spills in \map\ sub-task as 1.
By default, we used the input data of 4 GB size.

\begin{table}
\caption{Configurations of seven jobs.}
\label{t: job confs}
\centering
\begin{tabular}{ c|c|c|c|c|c|c|c }
\hline

Name      & Job$_1$ & Job$_2$ & Job$_3$ & Job$_4$ & Job$_5$ & Job$_6$ & Job$_7$ \\ \hline
cluster   & C1      & C1      & C1      & C1      & C2S     & C2H     & C2S     \\ 
\# map    & 1       & 2       & 3       & 4       & 6       & 6       & 2       \\ 
\# reduce & 2       & 2       & 2       & 2       & 2       & 2       & 2       \\ \hline

\end{tabular}
\end{table}

Other {\haa} optimization works (e.g.,~\cite{Shivanth:2011:PWC,Wang:2012:EMS}) used both TeraSort (TS) and WordCount (WC) as the benchmark job types in their evaluation.
In general, there is a difference in the output size between TS and WC types.
That is, the WC type is a lightweight I/O job, yielding a small output (i.e., ``word count") compared to the input,
while the TS type is a heavy I/O job, yielding almost the same size output from the input.
Despite the difference in their I/O request pattern,
however, the record processing time distribution of two types remain virtually identical (to be shown in Section~\ref{profiling_jobs}).
As such, by default, all seven jobs in Table~\ref{t: job confs} use the TS job type.

\subsection{Profiling {\haa} Jobs} \label{profiling_jobs}

{\haa} provides simple profile log data with the start/end time of sub-tasks, input/output data size of tasks, number of map/reduce tasks, etc.
Although another approach (\eg,~\cite{Tan:2010:VLC}) proposed a better tracing using only log information,
such a method is not directly usable to measure the quality of {\haa} optimization.
Starfish modified the Java profiler to extract additional information, such as the start/end time of sub-phases, for the {\haa} optimization at the cost of about 10--50\% overhead in profiling~\cite{Shivanth:2011:PWC}.
The benefit of this approach is that the modified Java profiler can run, independent of {\haa} version.

To analyze {\haa} tasks as proposed, however, we need to measure not only the type and start/end time of sub-phases, but also the start/end time of a ``record" processing (not available in the profiling approach used in Starfish).
Therefore, with these considerations,
in this paper, we instead modified {\haa} code directly to extract more detailed profiling data with a smaller profiling overhead.
Note that profiling cost for records is orders of magnitude more expensive than
that for sub-phases as there are much more number of records than sub-phases.
As such, we group a set of records into a unit and measure per unit
(empirically, we found that a unit of 5 records strikes a good balance between the cost and accuracy).

\begin{figure}
\centering
\includegraphics[width=0.50\columnwidth]{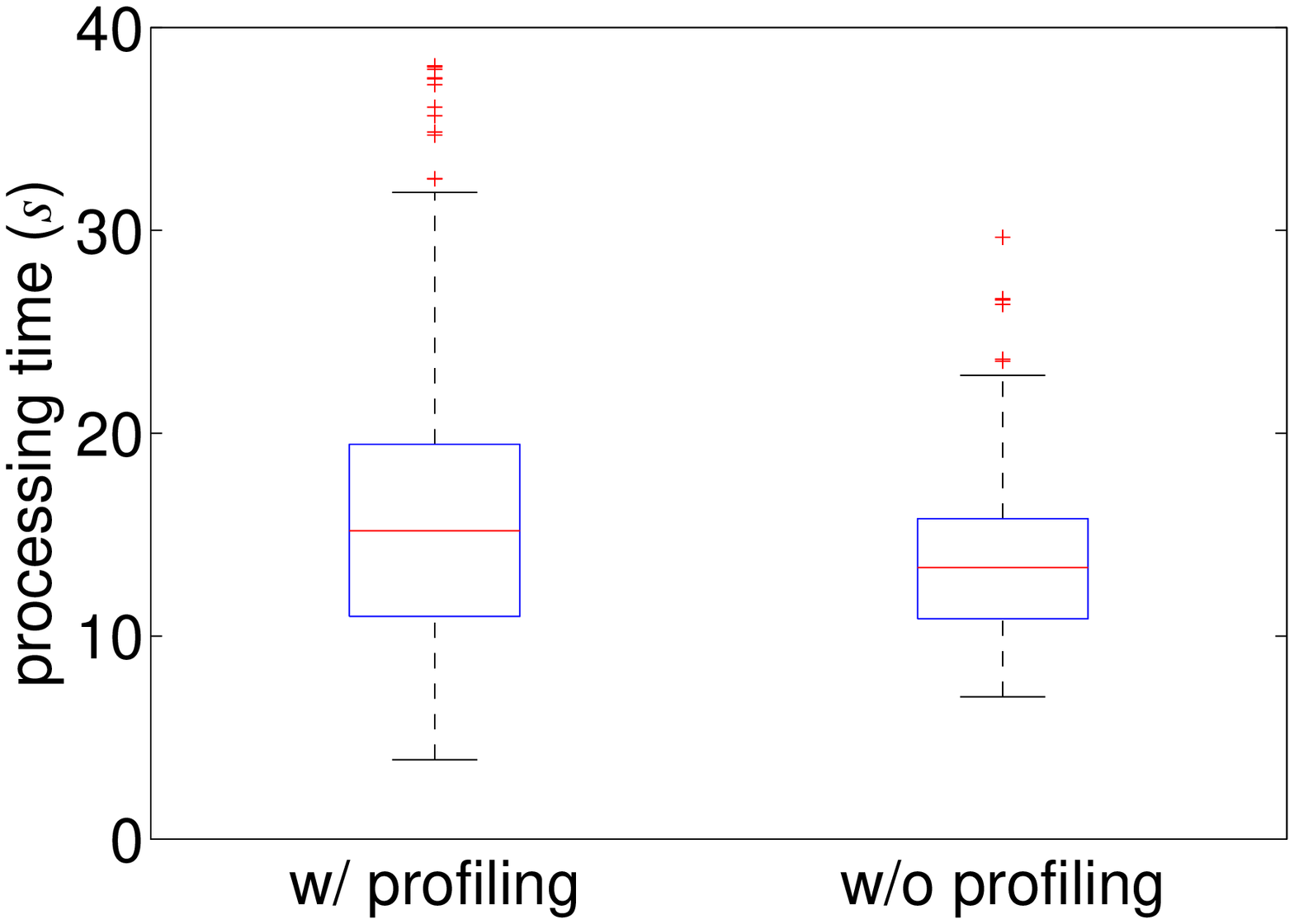}
\caption{Box plot of task processing times of two jobs: these box plots display difference between task processing times running with our profiler and without a profiler.}
\label{prof_over}
\end{figure}

As an illustration of the profiling overhead, for instance,
Figure~\ref{prof_over} shows the comparison of processing times of Job$_2$ between with and without profiling.
We used ten tasks closest to the median to get the average task processing time.
Note that the difference between the average map task processing time with and without profiling is from 4\% to 7\%, with the average of 5.3\%.
In order words, the overhead incurred from profiling is less significant than what was incurred in the Starfish profiler (i.e., 10--50\%),
despite the fact that our profiler collects more fine-grained profiling data than the Starfish profiler does.

\subsection{Distribution of Record Processing Time} \label{s:records dist}

\begin{figure}
\centering
  \begin{tabular}{cc}
    \hspace{-0.2in}
      \includegraphics[width=0.50\columnwidth]{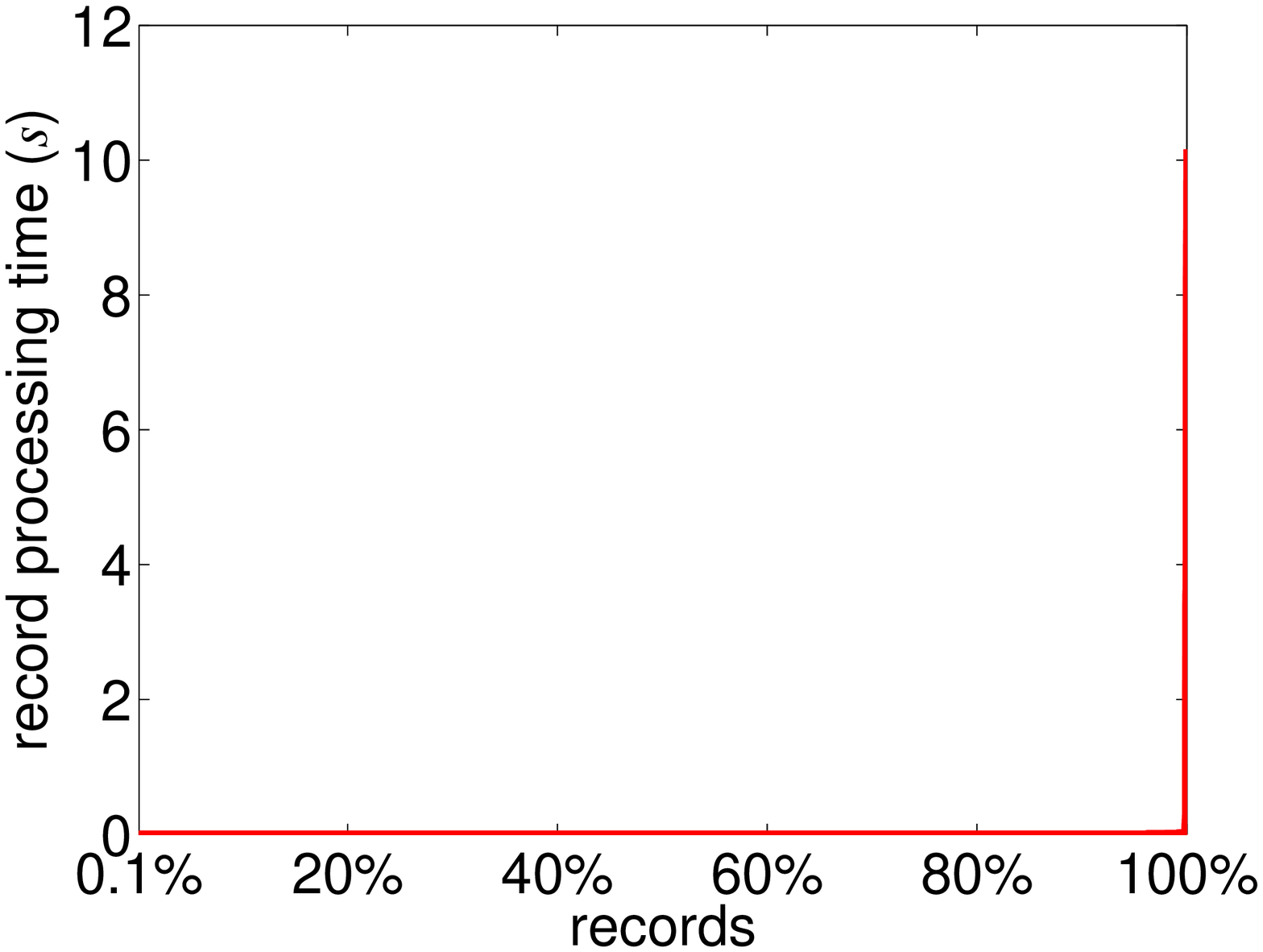} &
    \hspace{-0.2in}
      \includegraphics[width=0.50\columnwidth]{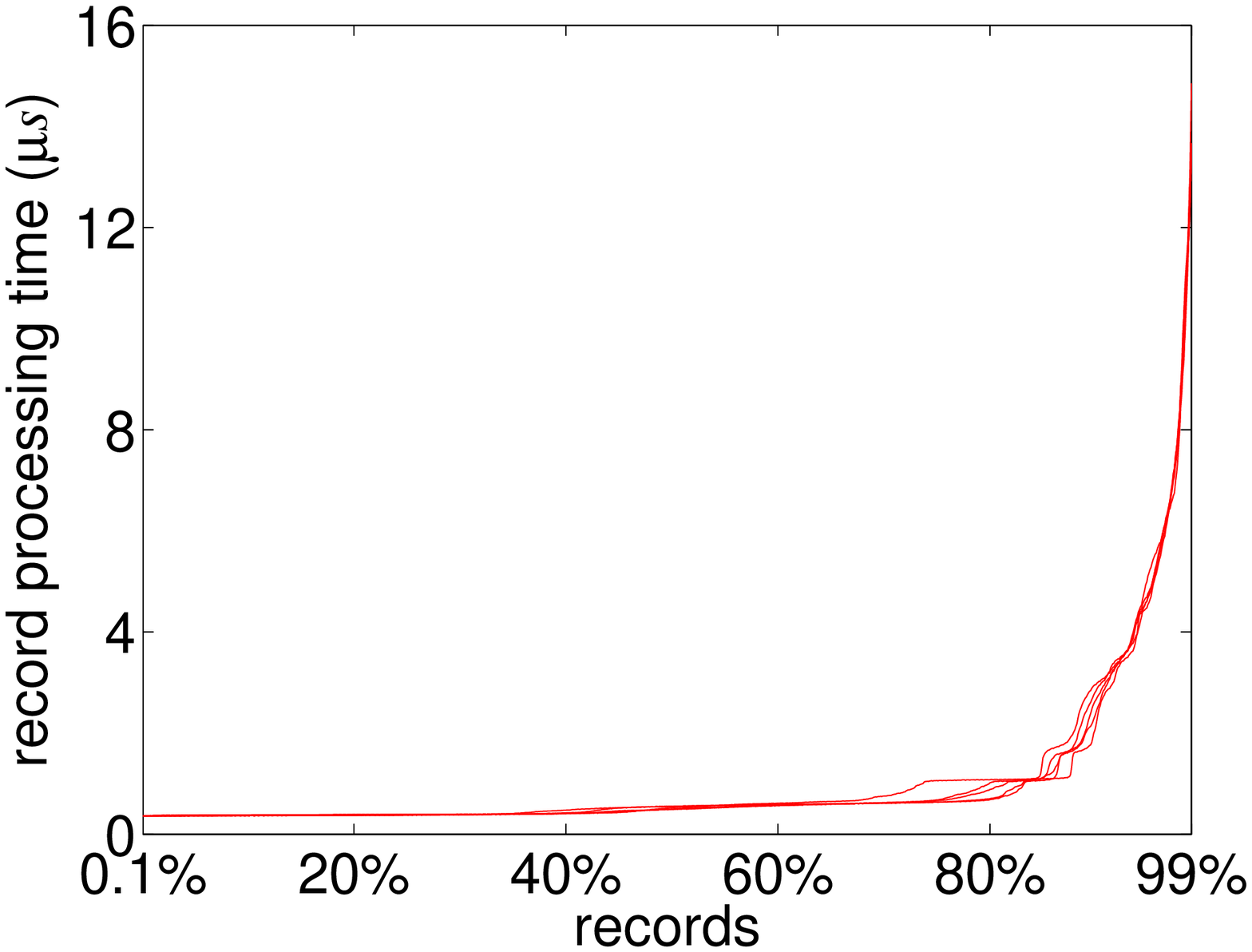} \\
      (a) With all records & (b) With 99\% records
  \end{tabular}
\caption{
The distribution of record processing times in the {\readmap} sub-phase of {\map} tasks (Job$_2$).
Each ``red" line represents a {\map} task.
All records are divided into 1,000 buckets by processing times.
$x$-axis is the ID of buckets and $y$-axis is the sum of all records processing time in each bucket.
In (b), 99\% records is used in order to enlarge the (a).}
\label{real_dist}
\end{figure}

We first analyze record processing times of different cases.
First, Figure~\ref{real_dist}(a) shows the distribution of record processing times in the {\readmap} sub-phase of Job$_2$.
Compared to the illustrated example in Figure~\ref{tasks},
the case with real data in Figure~\ref{real_dist} shows that the record processing times of a few records are several orders of magnitude longer than the rest of records.
Note that these a few records occupy the majority of processing times in the {\readmap} sub-phase.
Figure~\ref{real_dist}(b) shows an enlarged part of Figure~\ref{real_dist}(a),
showing only 0\%--99\% of records (excluding 1\% with the longest time).
Note also that the majority of records (e.g., about 85\% of records) takes very similar processing times.

An important characteristic of the distribution of record processing times is that
it is \textit{heavy-tailed}--i.e., a few records with a high values dominate the entire distribution.
The heavy-tailness refers to power tail distribution with tail index $\alpha >0$,
\begin{equation}
  \label{eq:heavy-tail}
  P(X > x) \sim c x^{-\alpha}
\end{equation}
with some positive constant $c >0$.
Heavy-tail phenomena is widely observed in many areas such as traffic data, hydrology, and finance
(see Resnick~\cite{resnick:2006} for more details).
Observe from (\ref{eq:heavy-tail}) that taking log-transformation gives
\[
  \log(1 - P(X >x) ) \sim \log c - \alpha \log x
\]
Hence, the heavy-tailed distribution can be assessed by plotting empirical distribution with respect to ordered data.
This plot is called \textit{emplot}, and intuitively speaking, the heavy-tailed distribution looks linear in log-log plot of the tail empirical distribution and ordered data.
Another popular method is to examine the value of heavy-tail index $\alpha$.
For example, the Hill estimator is defined as
\[
  \hat{\alpha}^H (k) = \frac{1}{k} \sum_{i=1}^k \left( \log Y_{n+1-i} - \log Y_{n-k} \right)
\]
where $Y_{i}$ is the $i$-th order statistics (or ($n+1-i$)-th largest or upper order statistics)
and $k$ is the number of upper-order statistics used in estimation.
The graph of $\{k, \hat{\alpha}^H (k) \}$ is called the {\em Hill} plot,
and a typical estimate of $\alpha$ is to use the stable one over many upper-order statistics used.

\begin{figure}
\centering
\includegraphics[width=1\columnwidth]{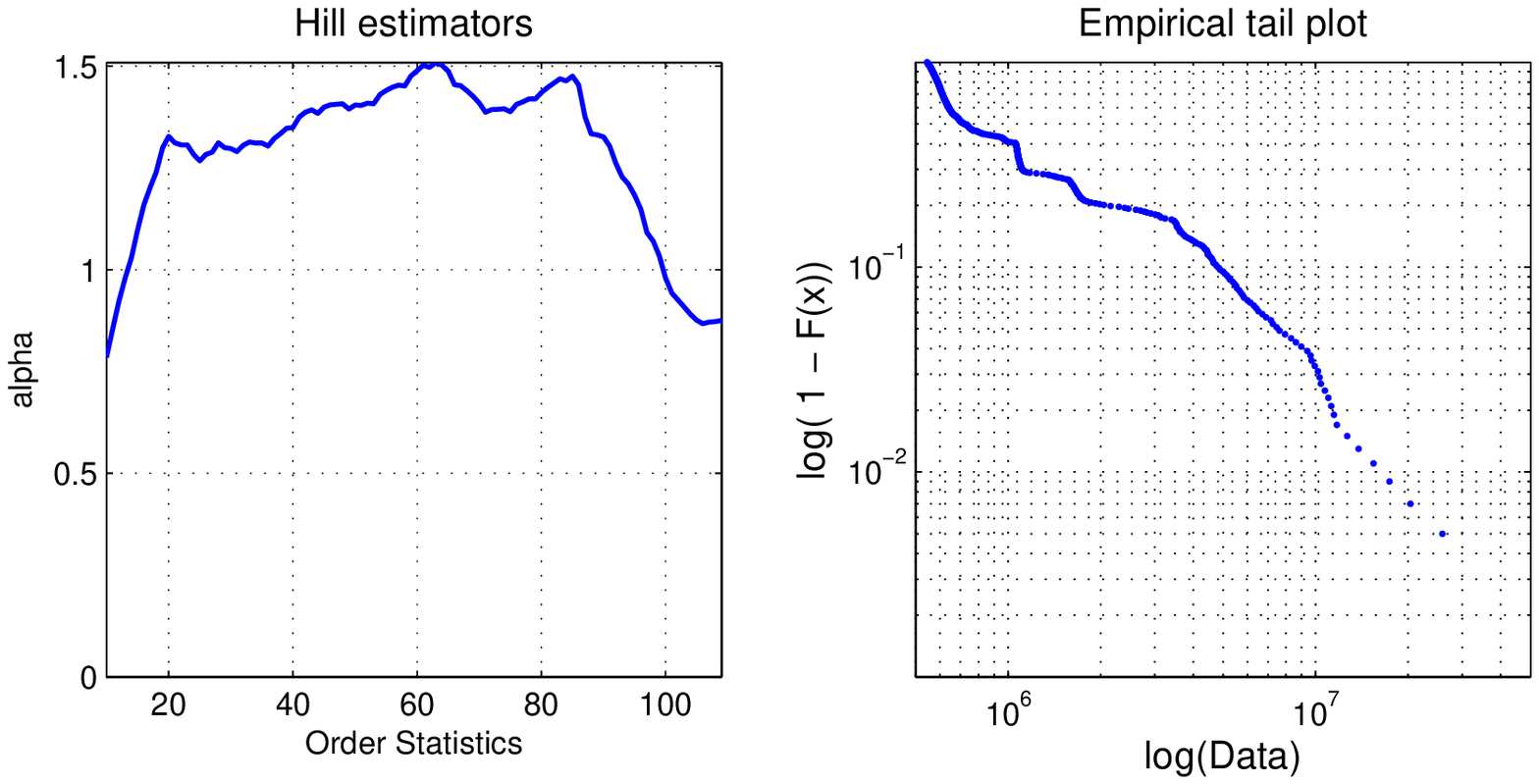}
\caption{The Hill plot and emplot  of Figure~\ref{real_dist}(a).}
\label{f:hill}
\end{figure}

Figure~\ref{f:hill} represents the Hill plot and emplot of Figure~\ref{real_dist}(a).
The tail index $\alpha$ is around 1.3 and emplot clearly shows the linearity between the tail empirical distribution and ordered data.
Therefore, it is concluded that the record processing times of {\readmap} sub-phase is heavy-tailed.
Note also that estimated tail index implies the finiteness of population mean, but infinite variance.
Indeed, record processing times are dominated by a few extremely large observations as we have observed in Figure~\ref{real_dist}.

\begin{figure}
\centering
  \begin{tabular}{cc}
    \hspace{-0.2in}
      \includegraphics[width=0.50\columnwidth]{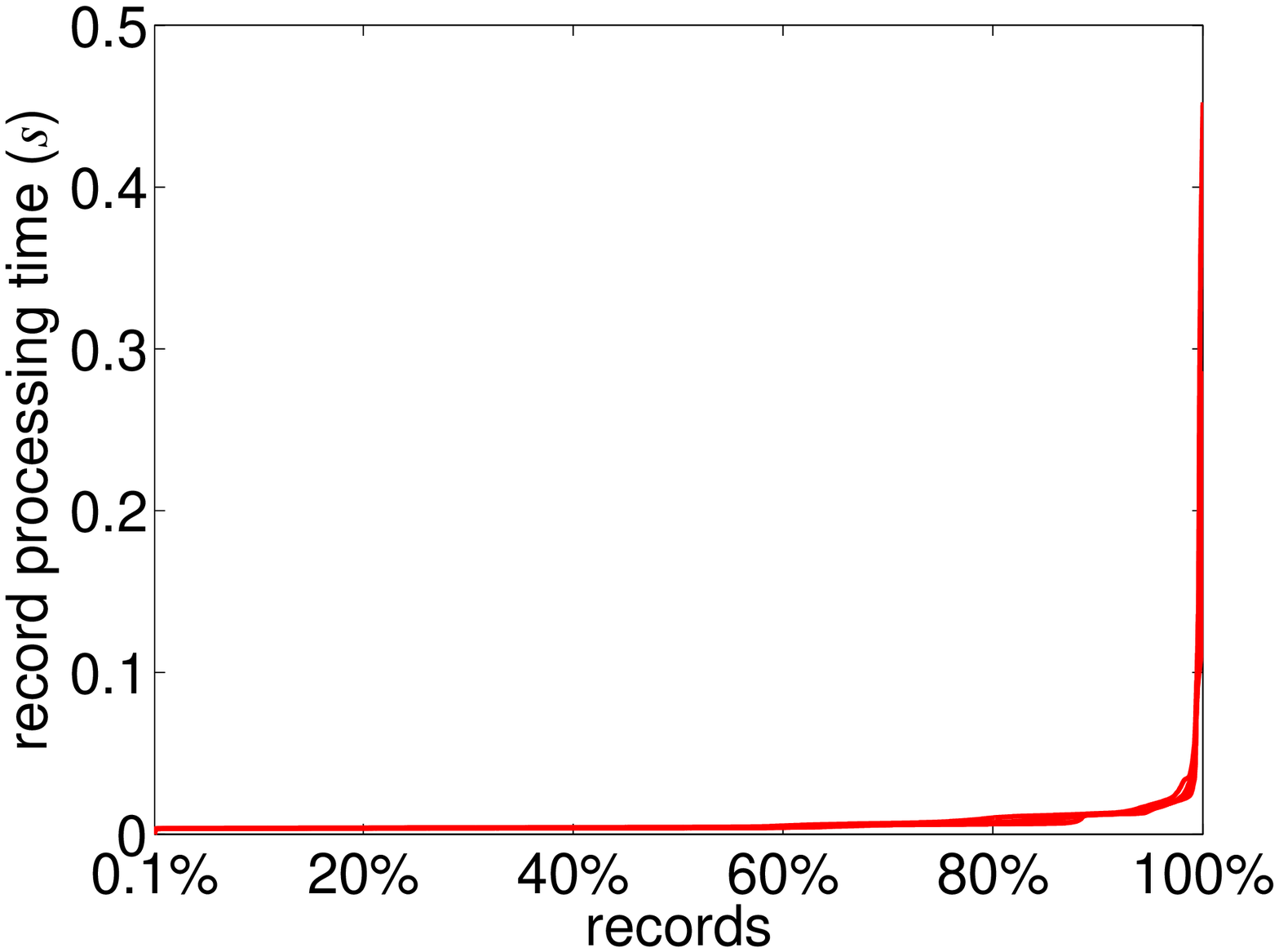} &
    \hspace{-0.2in}
      \includegraphics[width=0.50\columnwidth]{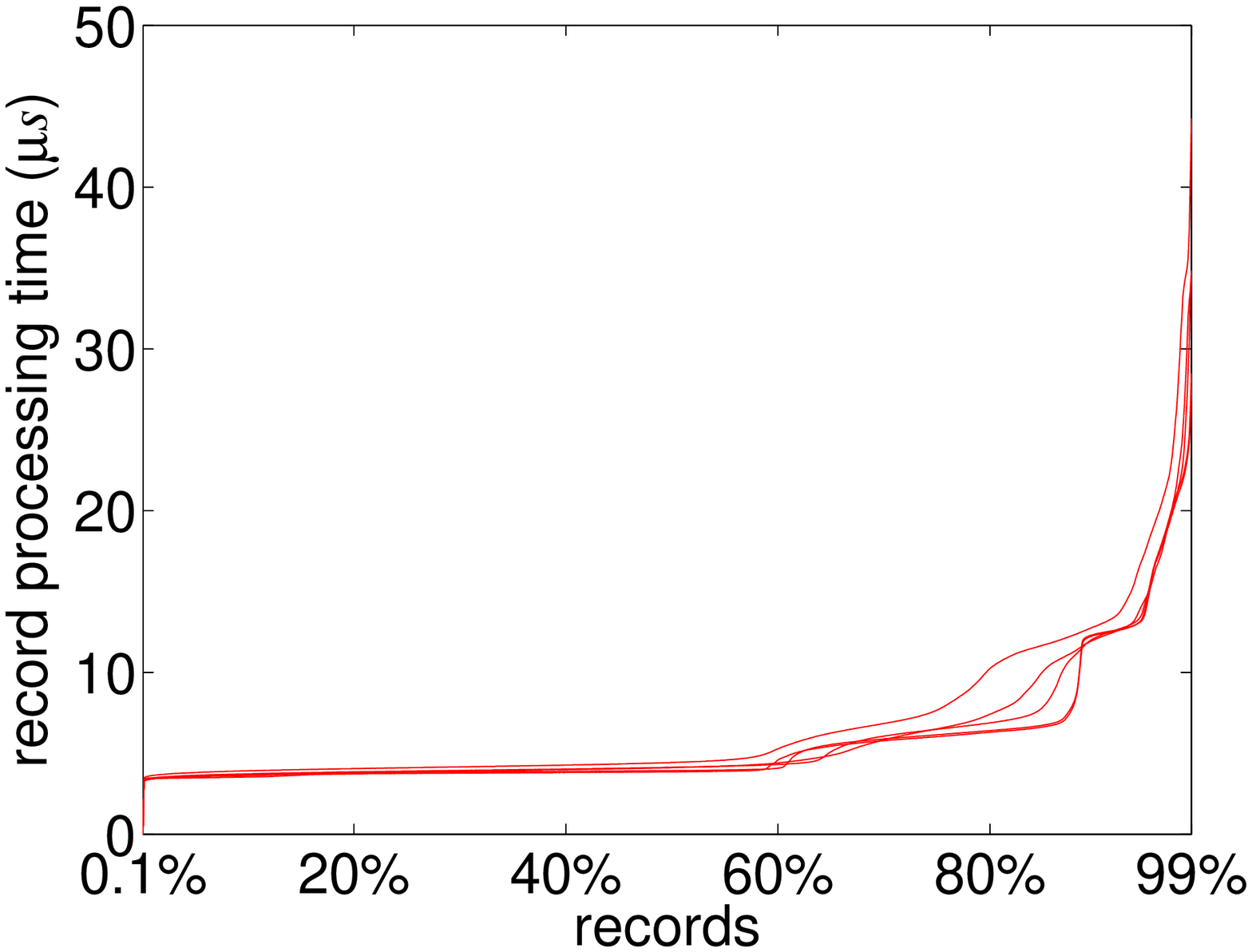} \\
      (a) With all records & (b) With 99\% records
  \end{tabular}
\caption{The distribution of record processing times in the {\reducewrite} sub-phase of {\reduce} tasks (Job$_2$).}
\label{comp_reduce}
\end{figure}

Second, in the case of {\reduce} task, the {\reducewrite} sub-phase is used to measure the quality of optimization.
Figure~\ref{comp_reduce} shows a similar pattern to Figure~\ref{real_dist}.
In general, input file sizes in {\reduce} task is uneven, unlike {\map} task.
The skewed file sizes cause a relatively less efficient performance in {\reduce} task than those in {\map} tasks.
A work such as~\cite{Ibrahim:2010:CCT} addressed this problem by using an efficient ``partitioning'' for more balanced distribution of input data in {\reduce} tasks.
Our proposed idea is not affected by file sizes and the number of records.
However, comparing the optimal cost of all {\reduce} tasks visually,
we used an evenly distributed data set and algorithm according to the number of reduce slots as a benchmark.

\begin{figure}[tb]
\centering
  \begin{tabular}{cc}
    \hspace{-0.2in}
      \includegraphics[width=0.50\columnwidth]{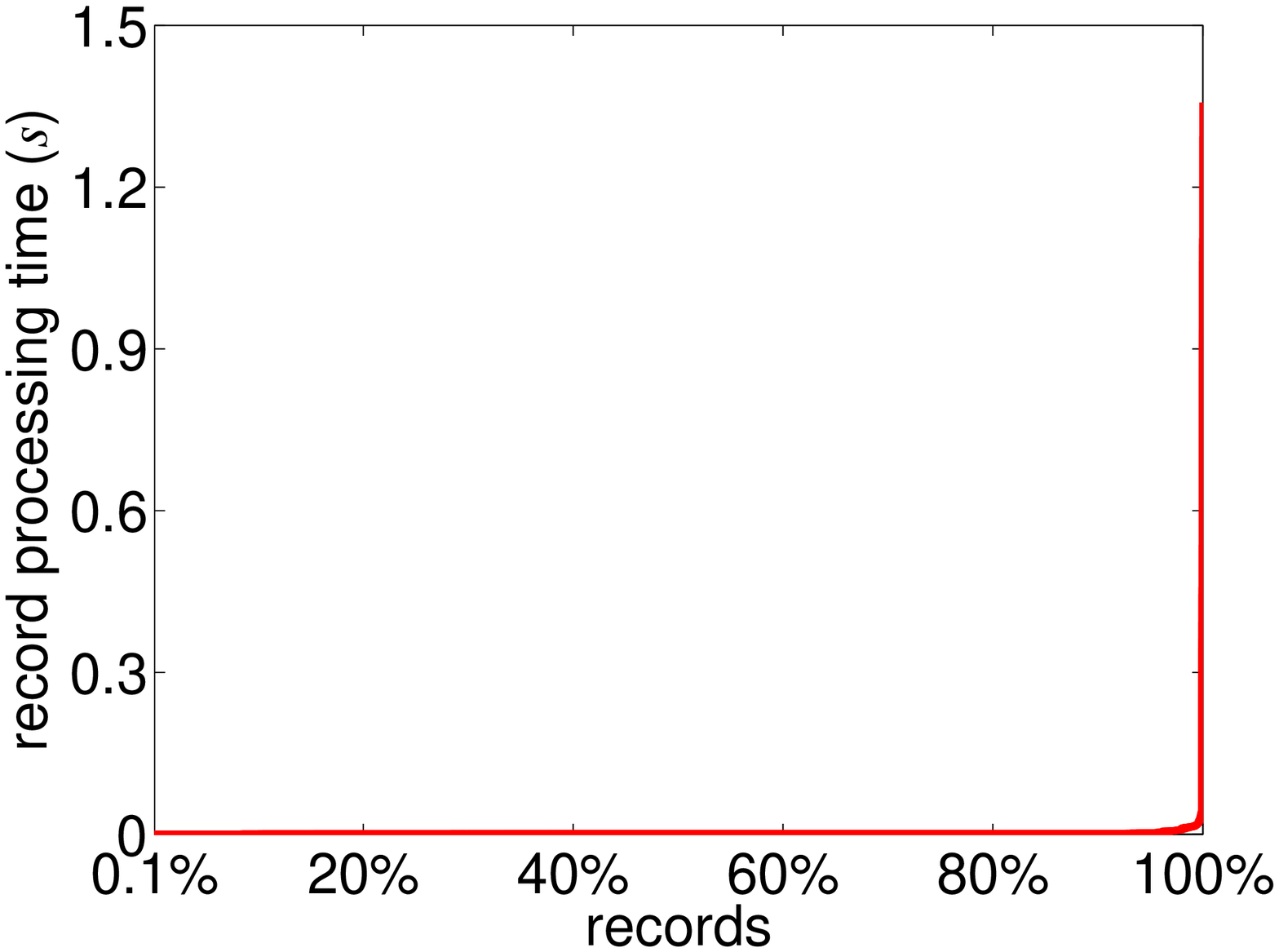} &
    \hspace{-0.2in}
      \includegraphics[width=0.50\columnwidth]{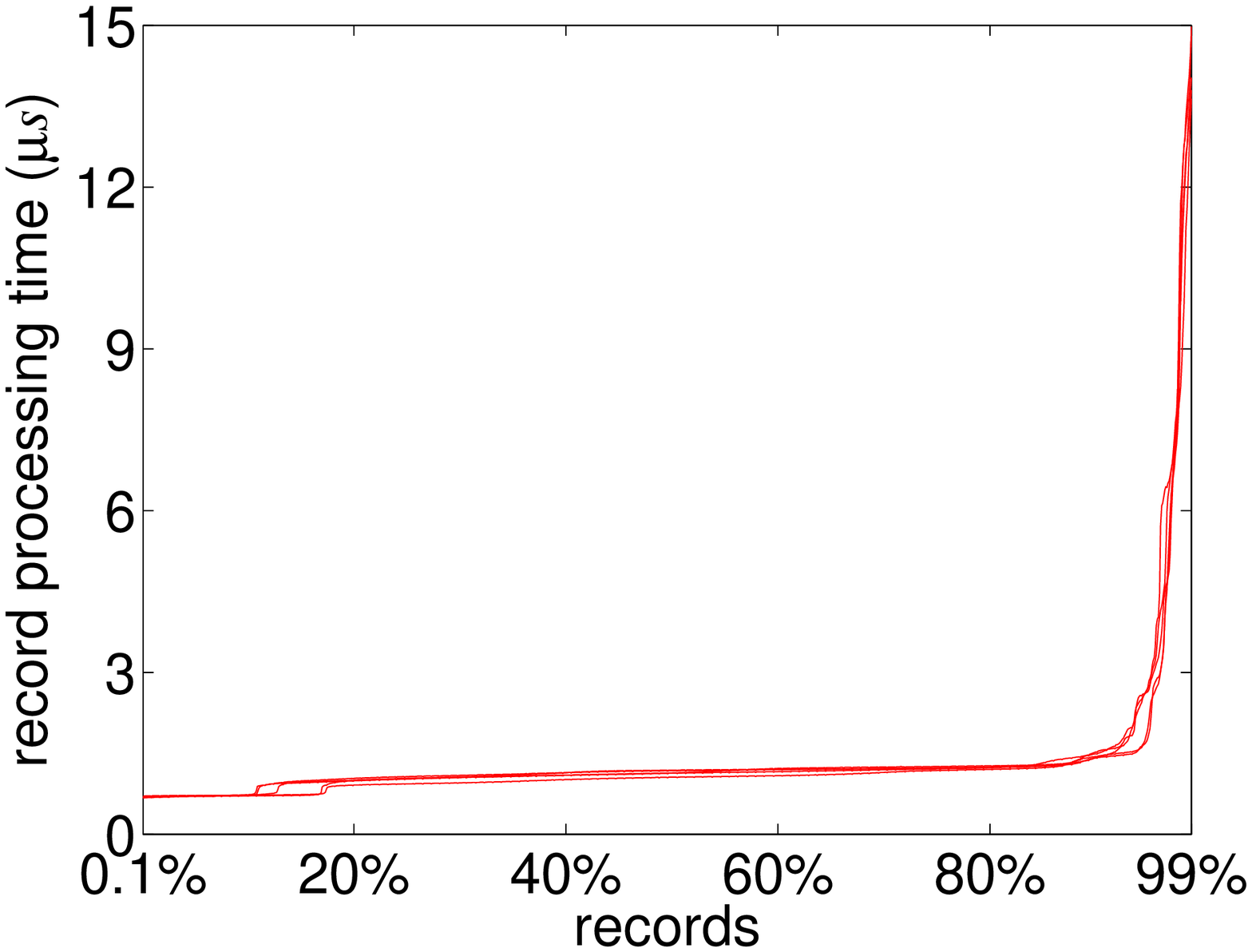} \\
      (a) With all records & (b) With 99\% records
  \end{tabular}
\caption{The distribution of record processing times in the {\readmap} sub-phase of {\map} tasks in a WordCount type job (as opposed to Job$_2$).}
\label{comp_jobtypes}
\end{figure}

Third, we compared the record processing times between two job types -- TeraSort (TS) and WordCount (WC).
Although the output data sizes of WC type jobs are smaller than those of TS type jobs,
there is no difference in their record processing times in {\readmap} sub-phases.
This is because the number of records to process is dependent on the input data sizes, not on the output data sizes.
The greatest benefit of having smaller output data sizes occurs in the {\spill} sub-phase when output is being written to HDD.
Therefore, the effect of output data size is minimal to our proposed measure.
Figure~\ref{comp_jobtypes} shows the distribution of processing times in the {\readmap} sub-phase of {\map} tasks in a WC type job, as opposed to Job$_2$ in Figure~\ref{real_dist}.
As explained, patterns between WC and TS type jobs remain virtually the same, and our measure is applicable to both types of jobs in {\haa}.
As such, for the remaining experiments, we only present jobs with TS type.


\subsection{Effectiveness of Our Proposal}

\begin{figure}[tb]
\centering
\includegraphics[width=0.49\columnwidth]{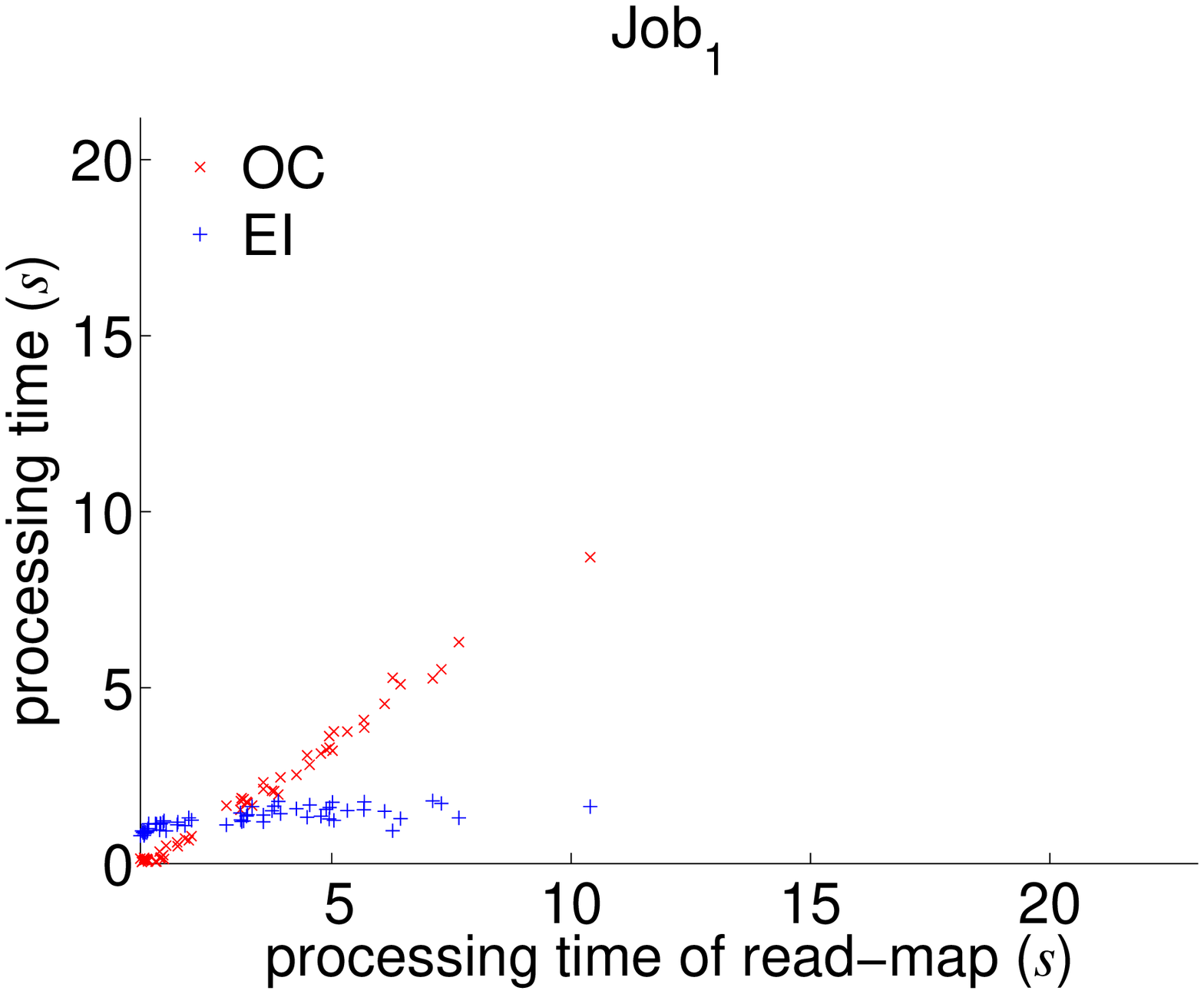}
\includegraphics[width=0.49\columnwidth]{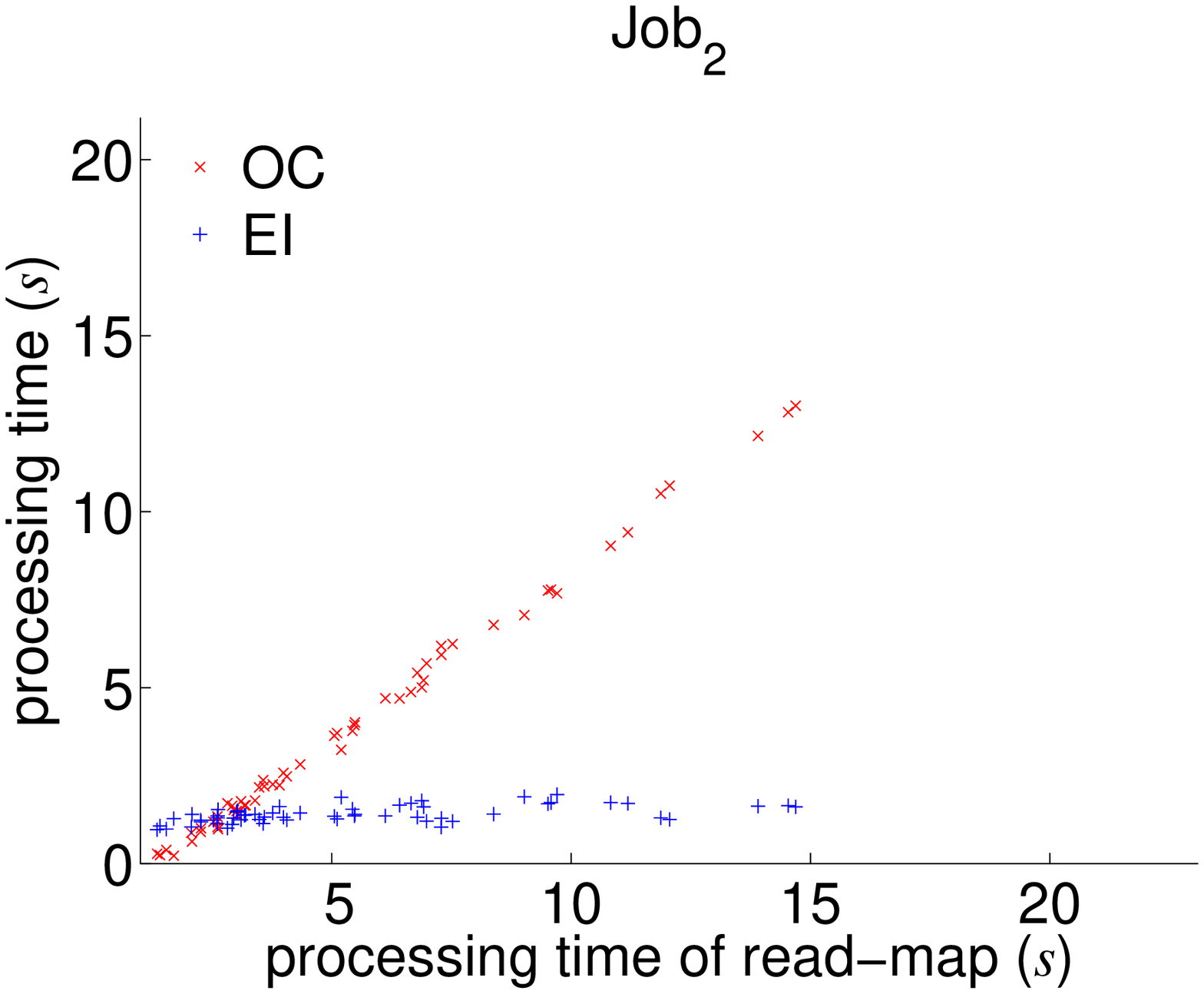}
\includegraphics[width=0.49\columnwidth]{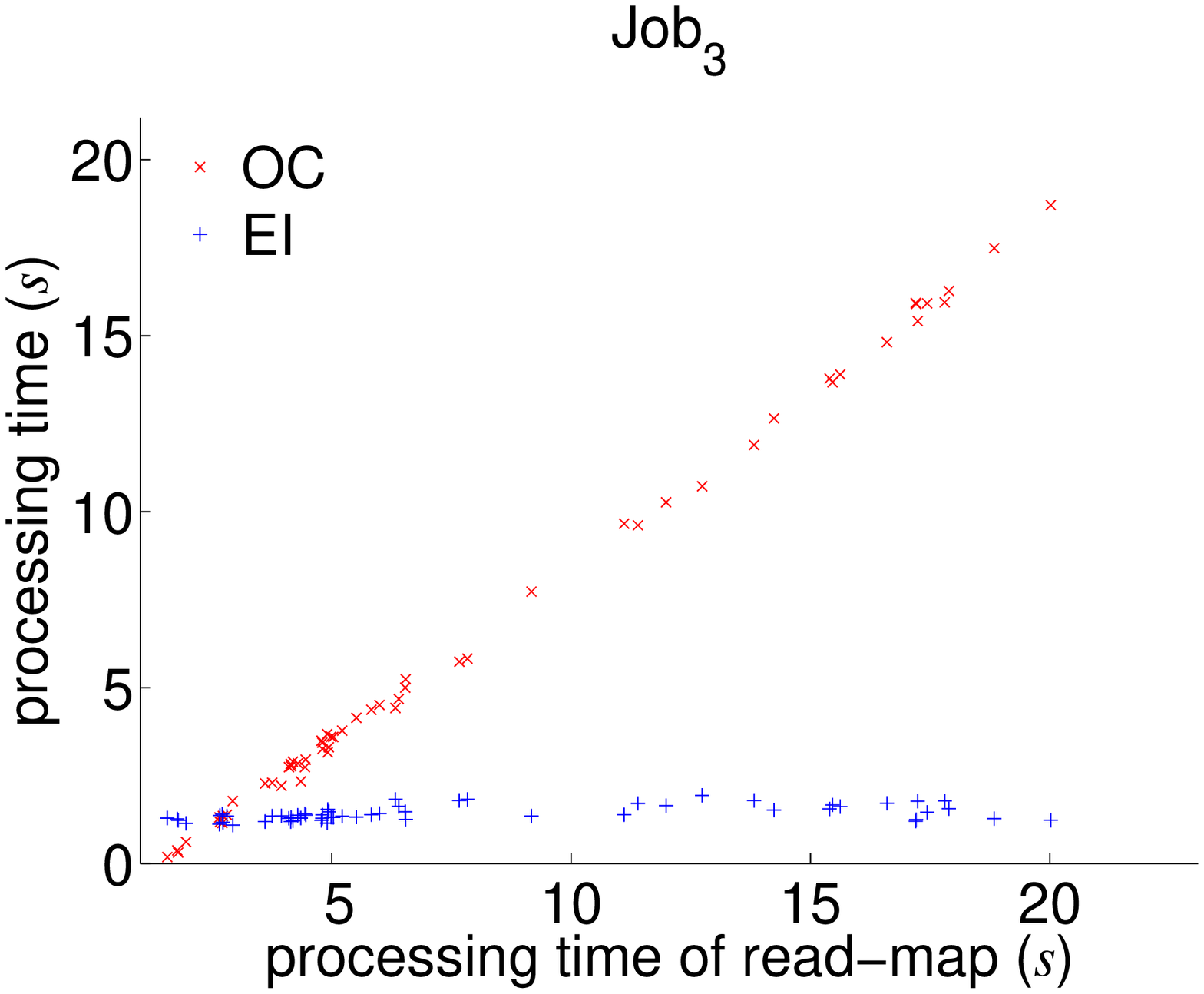}
\includegraphics[width=0.49\columnwidth]{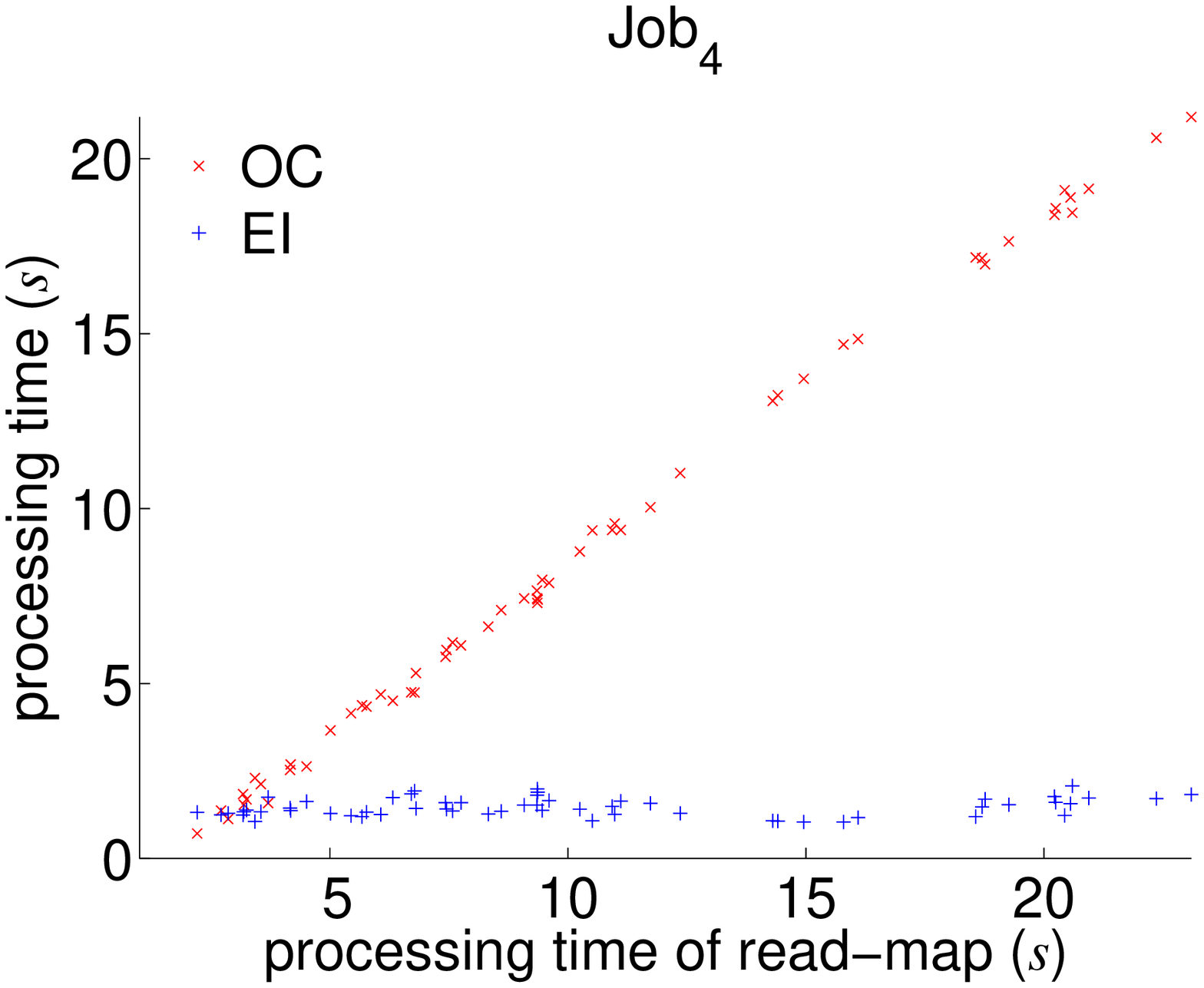}
\caption{Relationship between $OC$ and $EI$ with varying number of map slots.}
\label{f: change map slots}
\end{figure}

So far, we have shown that our inference about the distribution of record processing time in Section~\ref{s:opt-measure} is correct.
In this section, we evaluate the goodness of $EI$ (\ie, estimated ideal cost) which is defined in Section~\ref{s:stat-analysis}.
An important property of $EI$ is its ``consistency."
That is, any {\map} tasks of the same job type running on the same H/W (\ie, CPU, Disk) should exhibit very similar $EI$ regardless of their H/W utilization.
To demonstrate the consistency of $EI$, we show that tasks have similar $EI$ scores, even if they have varying record processing times.
Figure~\ref{f: change map slots} shows the relationship between the processing times of {\readmap} sub-phase and $EI$ and $OC$.
Table~\ref{t: change map slots} shows the summary of the results.
In order to compare the difference between the graphs visually, $x$-axis and $y$-axis are fixed to the same range.
$Job_1$, $Job_2$, $Job_3$, and $Job_4$ run on the same cluster, C1, but with different usable H/W resources.
For instance, $Job_1$ could use much more H/W resources (if required) than $Job_4$ could because the number of map slot in $Job_1$ is 1 (as opposed to 4 in $Job_4$).
As a result, the average processing time of {\map} tasks in $Job_1$ is much faster than that in $Job_4$.
Nevertheless, note that $EI$ scores remain constant.
For example, in Table~\ref{t: change map slots}, when $PR$ scores change from 3.219s in $Job_1$ to 10.309s in $Job_4$, $EI$ scores change from 1.259s to 1.453s, respectively.
That is, our proposed $EI$ is less affected by H/W utilization, showing a good consistency.
In addition, the standard deviations of $EI$ scores are much smaller (\eg, 0.211s--0.79s) than those of $PR$ (\eg, 2.143s--6.147s).

\begin{table}
\caption{Statistics of $PR$ and $EI$ of four jobs with varying number of map slots.}
\label{t: change map slots}
\centering
\begin{tabular}{c|c|c|c|c|c}
\hline
\multicolumn{2}{c|}{Type} & $Job_1$ & $Job_2$ & $Job_3$ & $Job_4$ \\
\hline
\multicolumn{2}{c|}{number of map slots} & 1 & 2 & 3 & 4 \\ \hline
\multirow{2}{*}{$PR$} & mean & 3.219$s$ & 5.441$s$ & 7.862$s$ & 10.309$s$ \\
                      & std  & 2.143$s$ & 3.525$s$ & 5.609$s$ & ~6.147$s$ \\ \hline
\multirow{2}{*}{$EI$} & mean & 1.259$s$ & 1.391$s$ & 1.415$s$ & ~1.453$s$ \\
                      & std  & 0.279$s$ & 0.241$s$ & 0.211$s$ & ~0.259$s$ \\ \hline
\multicolumn{2}{c|}{$vet_{job}$} & 2.407 & 3.790  & 5.413    & 7.190     \\ \hline
\end{tabular}
\end{table}

In Table~\ref{t: change map slots}, we obtained varying scores of $vet_{job}$, ranging from 2.407 to 7.19, depending on the configurations of jobs.
Recall that when $vet_{job}=1$, in theory there is no overhead cost and we achieve the ideal optimization.
Therefore, all four jobs clearly indicate that their current configuration is {\em not} yet optimal.
As how to ``automatically" find a better configuration to achieve $vet_{job}=1$ is beyond the scope of this paper and left for future work,
here we demonstrate that it is indeed possible to improve the configuration of those jobs ``manually" for the improved $vet_{job}$ score.

\begin{figure}[tb]
\centering
  \begin{tabular}{cc}
    \hspace{-0.2in}
      \includegraphics[width=0.49\columnwidth]{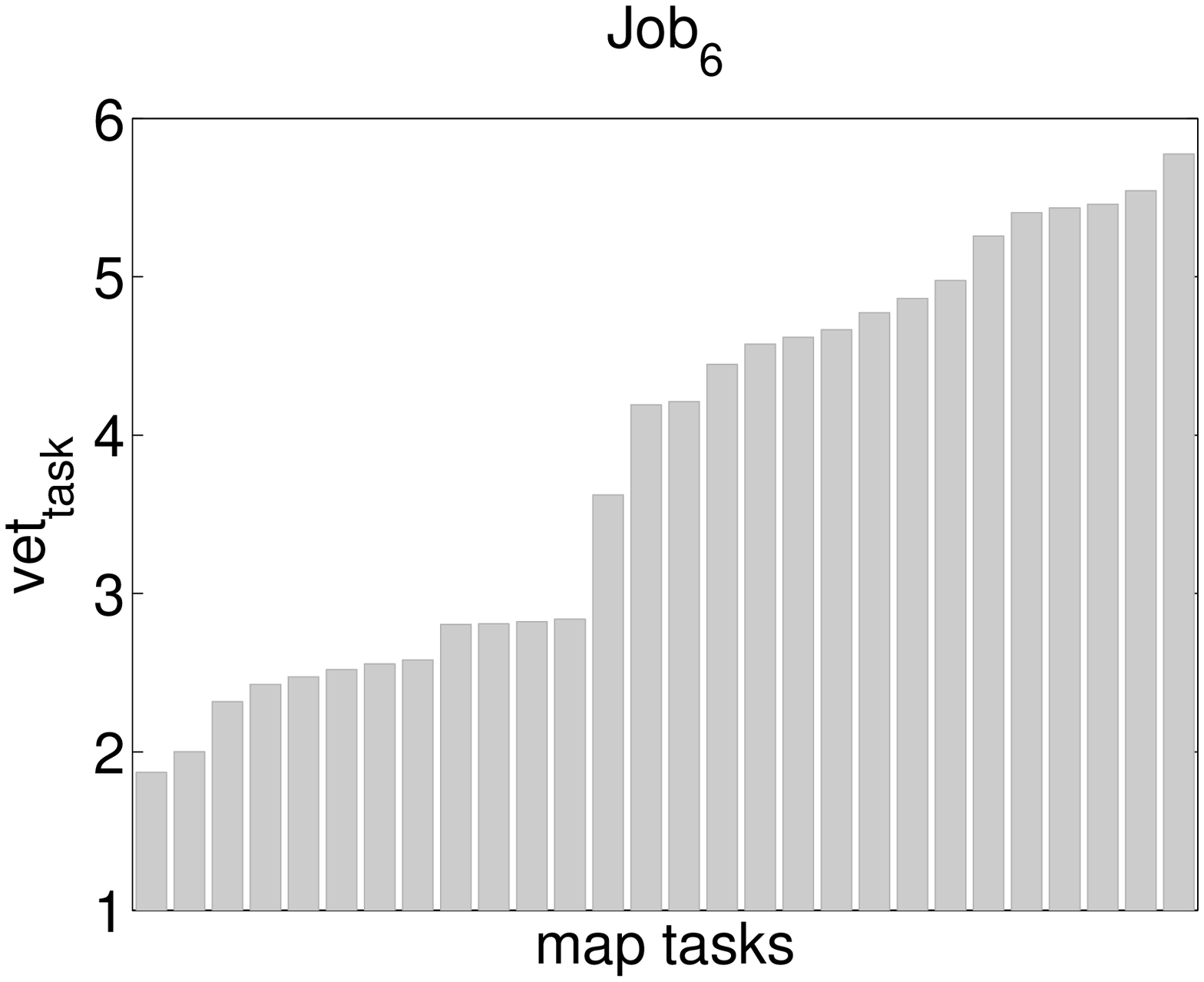} &
    \hspace{-0.2in}
      \includegraphics[width=0.5\columnwidth]{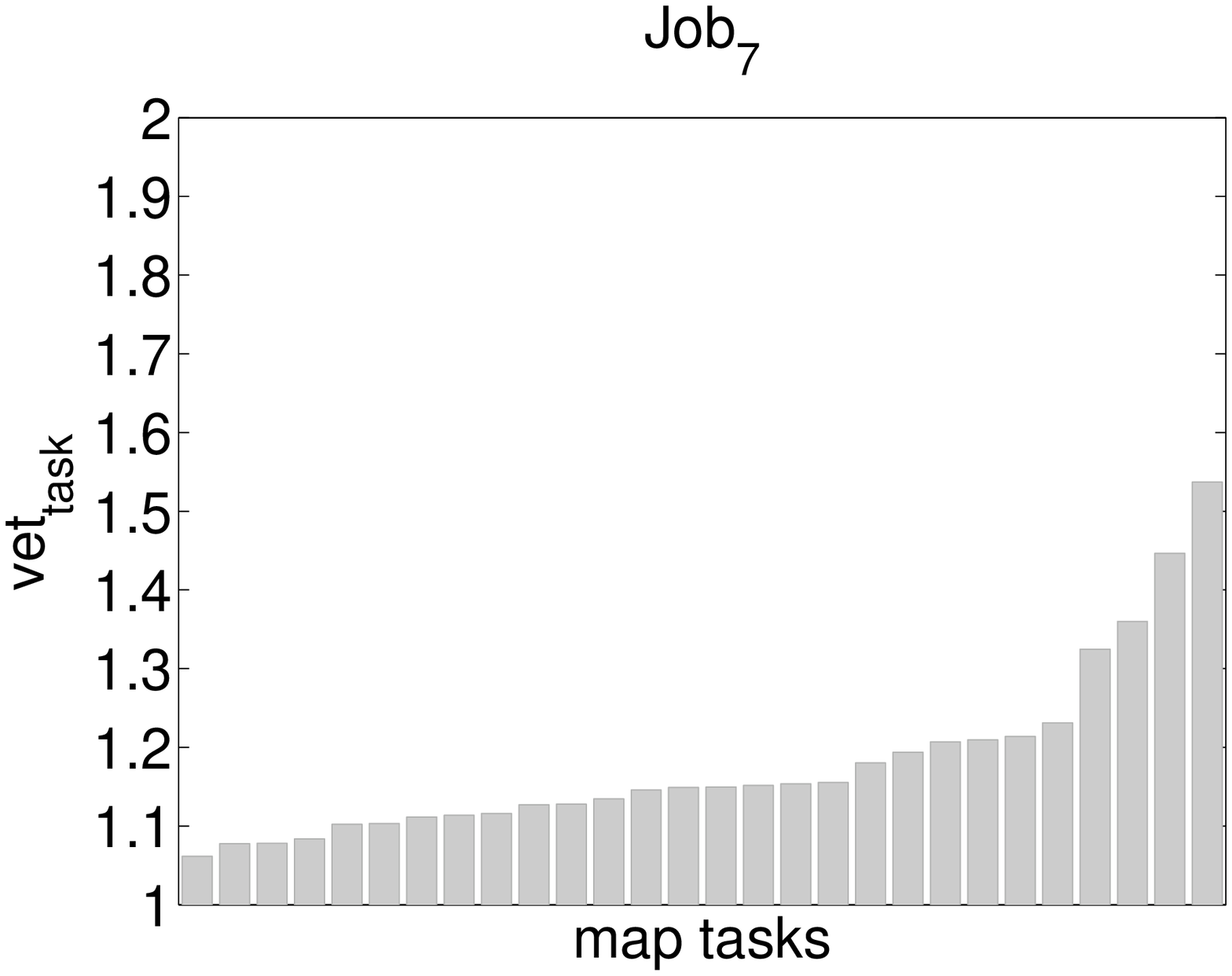} \\
      (a) Using HDD & (b) Using SSD
  \end{tabular}
\caption{Comparing $vet_{task}$s of jobs using HDD and SSD.}
\label{f: ideal vet task}
\end{figure}

Ideally, when H/W resources fully support a running job, its $vet_{job}$ approaches 1.
Figure~\ref{f: ideal vet task} shows a comparison between $Job_7$ using SSD and $Job_6$ using HDD.
The faster SSD would incur less CPU overhead.
Note that $vet_{task}$ of $Job_7$ is substantially improved from $vet_{task}$ of $Job_6$.
That is, the majority of $vet_{task}$ scores in $Job_7$ using SSD are around 1.3, while all $vet_{task}$ scores in $Job_6$ are no smaller than 1.9.

\begin{figure}[tb]
\centering
\includegraphics[width=0.49\columnwidth]{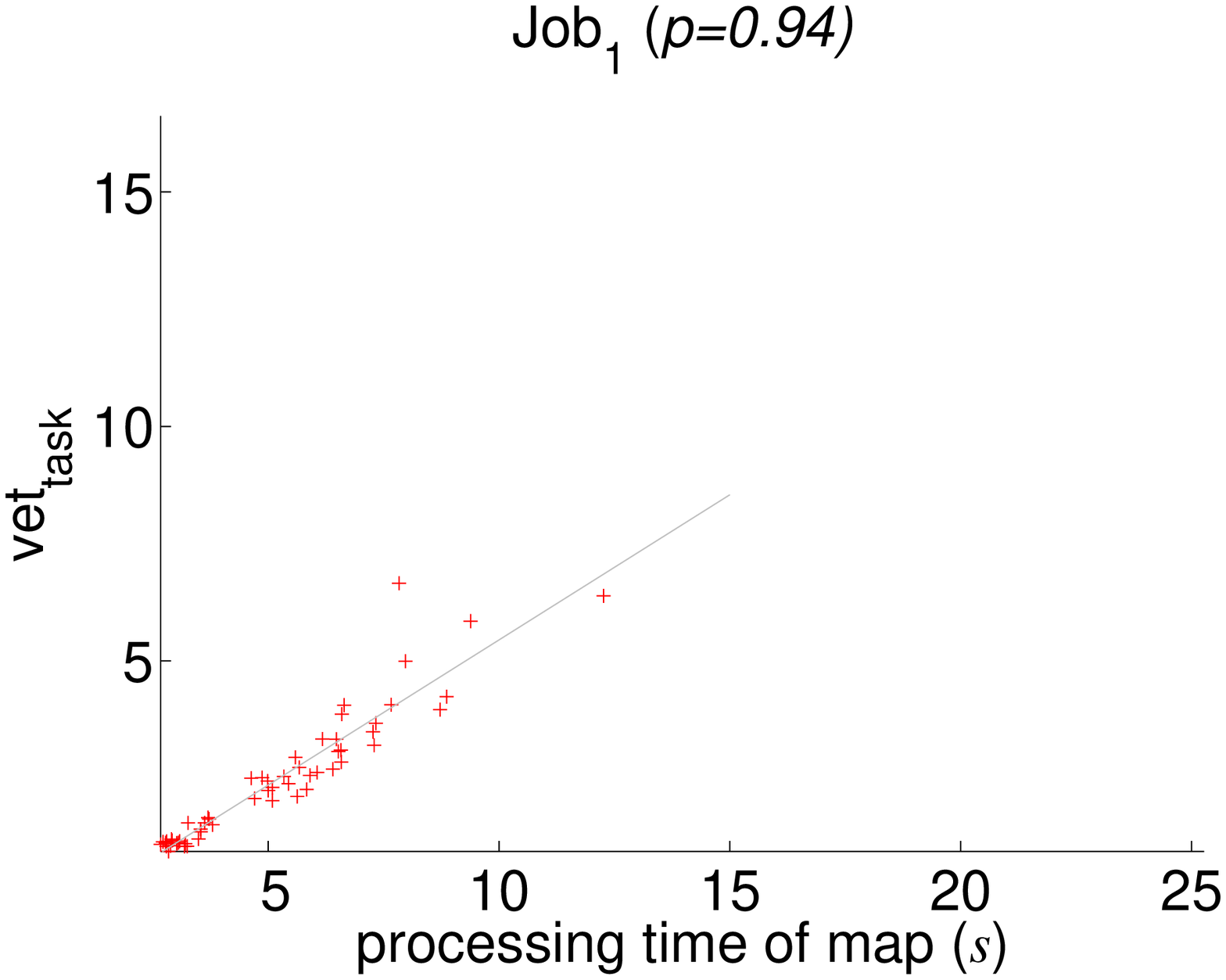}
\includegraphics[width=0.49\columnwidth]{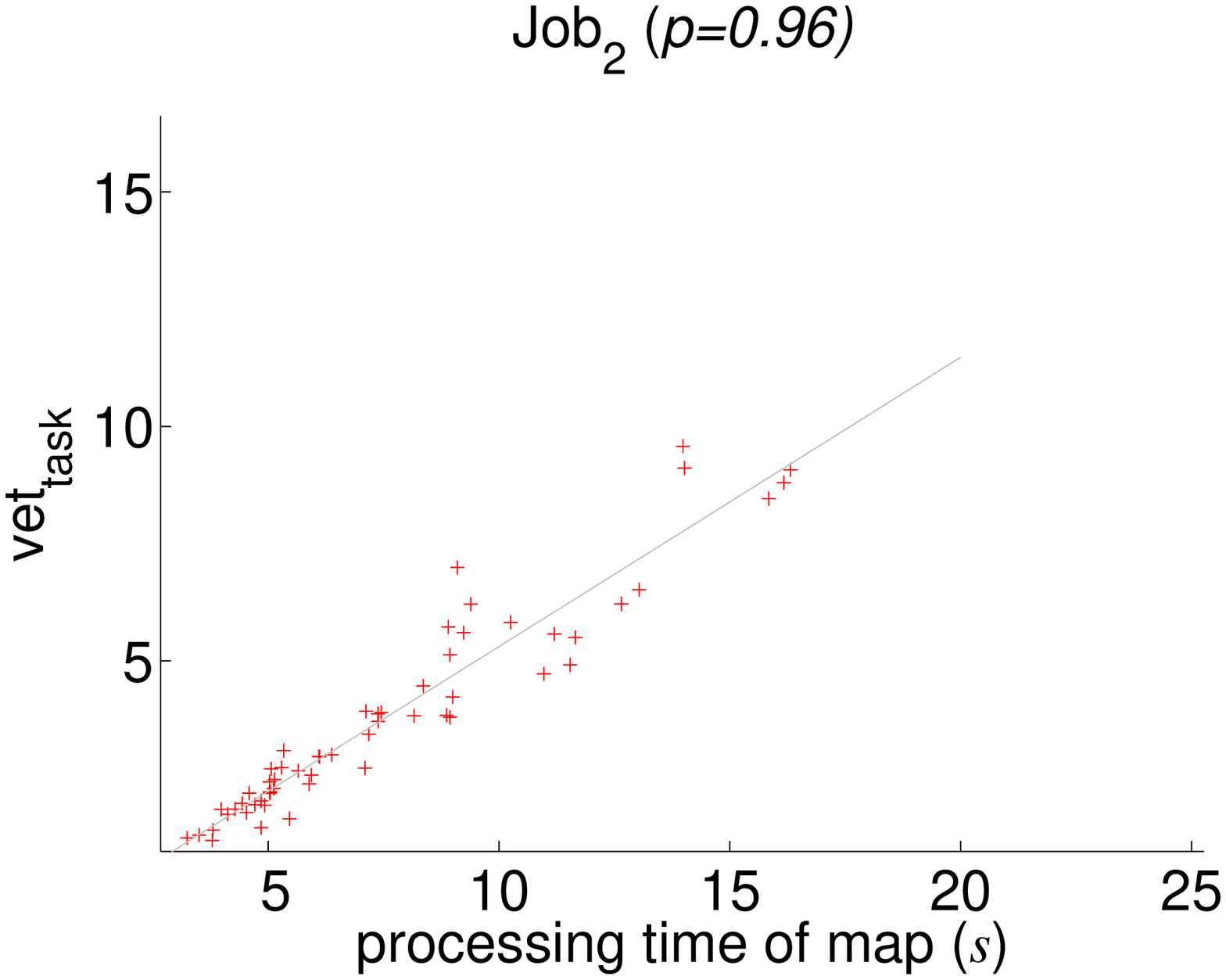}
\includegraphics[width=0.49\columnwidth]{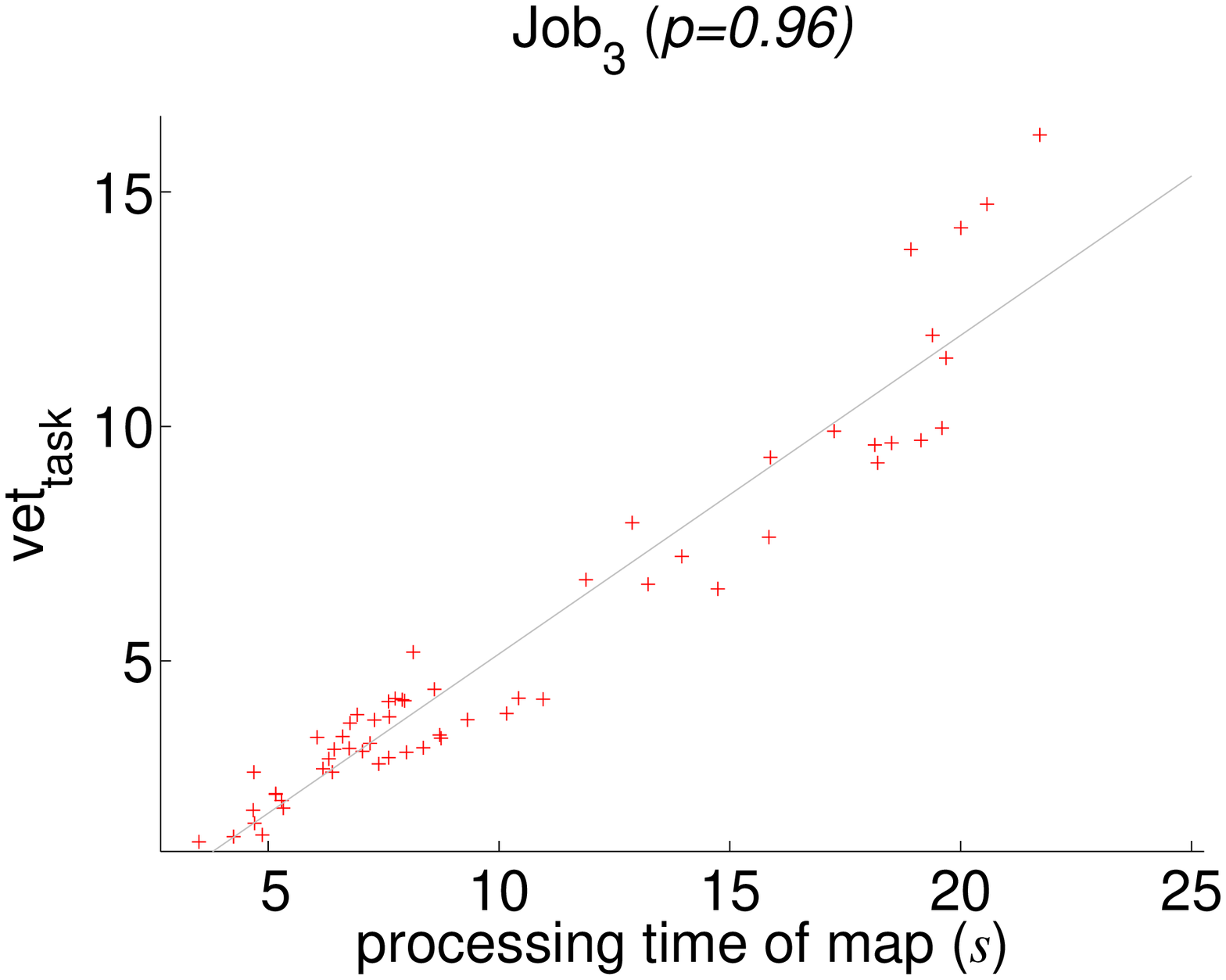}
\includegraphics[width=0.49\columnwidth]{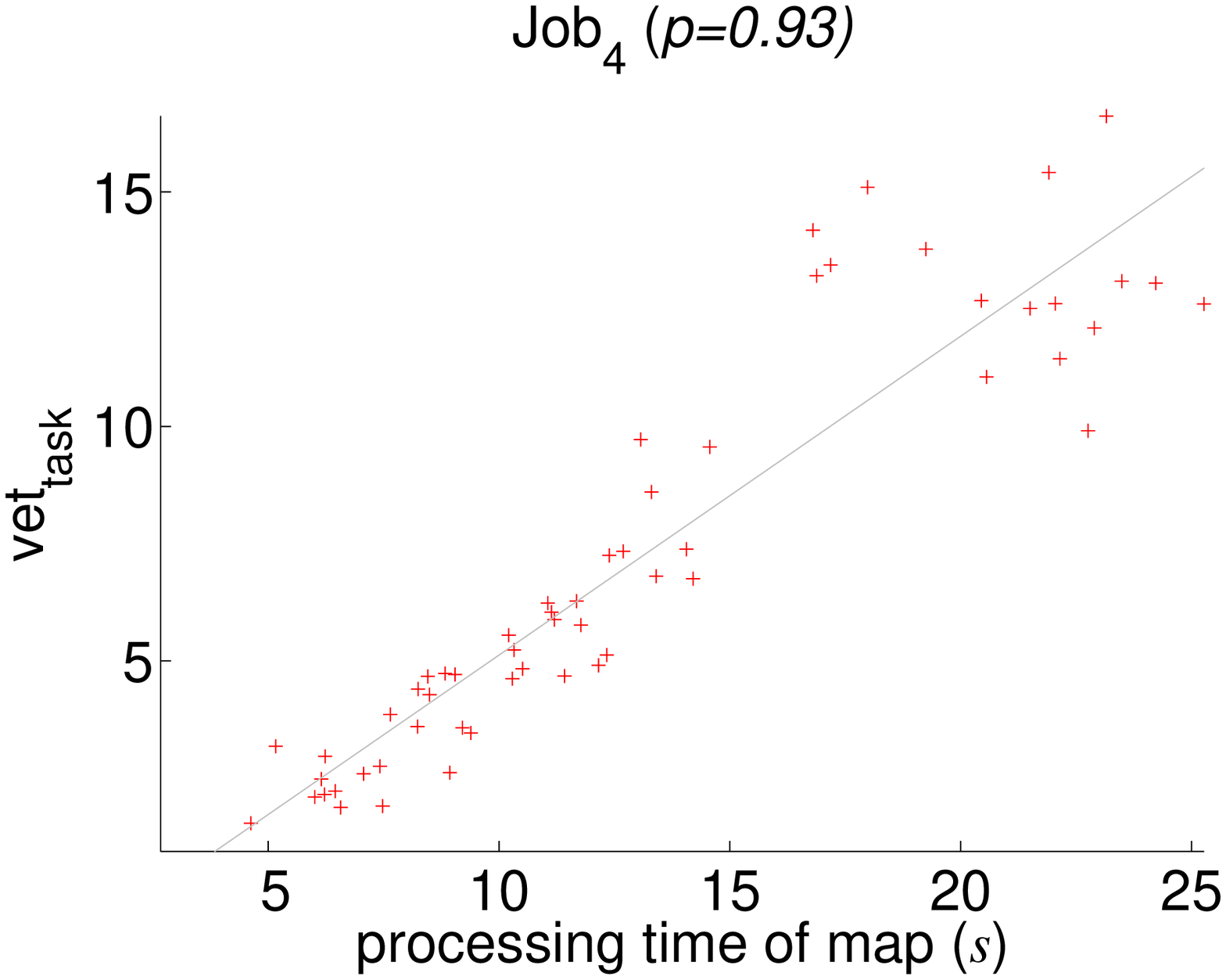}
\caption{$vet_{task}$ and the processing time of {\map} tasks: $p$ in title refers to the Pearson correlation score.}
\label{f: correlation vet and task}
\end{figure}

In addition, Figure~\ref{f: correlation vet and task} shows the scatter plots of $vet_{task}$ and the processing time of {\map} tasks.
Their high correlation is clearly visible (\ie, Pearson correlation scores of $Job_1$, $Job_2$, $Job_3$ and $Job_4$ are 0.94, 0.96, 0.96 and 0.93, respectively).
Therefore, our proposed measure of $vet_{task}$ is a good indication of the performance of {\map} tasks in {\haa}.
In conclusion, we show that it is indeed possible to further improve the configuration of jobs in {\haa} for a better performance, although we tune it manually,
and our proposed idea to measure the optimization of {\haa} by means of EI and $vet_{task}$ is effective.

\subsection{Complementing Starfish}

Starfish is a well-known {\haa} optimization framework.
Starfish optimizes {\haa} jobs by locating a parameter set, minimizing the cost model and often the predicted processing times of {\map} and {\reduce} tasks.
In this section, we show how to use our proposed measures to further help the optimization by Starfish.
Using Starfish \textit{version} 0.3.0, we ran four TS type jobs, $Job_a$, $Job_b$, $Job_c$ and $Job_d$, on C1, each using different parameter sets, suggested by Starfish.
All jobs use the number of {\map} and {\reduce} slots as 2 and 2, respectively.
Note that the ideal number of {\spill} sub-phase in {\map} tasks is 1, also correctly (but implicitly) identified by Starfish.
In Table~\ref{t: starfish}, the $PR$ scores of four jobs, although run on the same cluster with the optimized parameter sets that were suggested by Starfish, vary widely, ranging from 3.819s to 6.335s.
However, the $EI$ scores remain constant across four jobs and are very similar to the $EI$ score of $Job_2$ in Table~\ref{t: change map slots}, where we manually configured the parameter set.
This result suggests that although Starfish does a good job in finding near-optimal parameter sets for the given job,
there is still room for improvement (\ie, reducing $vet_{task}$ score to 1).


\begin{table}
\caption{Statistics of $PR$ and $EI$ of jobs optimized by Starfish.}
\label{t: starfish}
\centering
\begin{tabular}{c|c|c|c|c|c}
\hline
\multicolumn{2}{c|}{Type} & $Job_a$ & $Job_b$ & $Job_c$ & $Job_d$ \\
\hline
\multicolumn{2}{c|}{number of map slots} & 2 & 2 & 2 & 2 \\ \hline
\multirow{2}{*}{$PR$} & mean & 5.801$s$ & 3.819$s$ & 6.335$s$ & 4.189$s$ \\
                      & std  & 3.295$s$ & 2.364$s$ & 3.974$s$ & 2.244$s$ \\ \hline
\multirow{2}{*}{$EI$} & mean & 1.412$s$ & 1.387$s$ & 1.417$s$ & 1.364$s$ \\
                      & std  & 0.210$s$ & 0.206$s$ & 0.259$s$ & 0.235$s$ \\ \hline
\multicolumn{2}{c|}{$vet_{job}$} & 4.028 & 3.269 & 4.179 & 3.581  \\ \hline
\end{tabular}
\end{table}

Eventually, if one uses a H/W resource aware scheduler that can dynamically control the number of tasks
(as opposed to the fixed number as the sum of the number of {\map} and {\reduce} slots),
then a further optimization than that currently supported by Starfish can be algorithmically achieved.
Recently, for instance, works such as~\cite{Yong:2009:TRA,Wolf:2010:FAA,Polo:2011:RAA} investigated resource aware schedulers for {\haa}.
In such a case, our proposed measure, $vet_{task}$, will be helpful for the scheduler to select the number of tasks.
For instance, given the number of tasks calculated as 4, if the $vet_{task}$ of the tasks is higher than 4,
the scheduler should reduce the number of tasks.
We plan to pursue the resource aware scheduler used together with the proposed $vet_{task}$ in future.

\section{Related Works}

Existing research have used different types of measurement to show the optimality of a \ha optimization method, either directly or indirectly.
Table~\ref{t:measures_optimization} summarizes three types of measurements used in literature.
\begin{table}
  \caption{Summary of measures used for the \ha optimization}
  \label{t:measures_optimization}
  \centering
  \begin{tabular}{ c p{0.60\columnwidth} }
  \hline
  Type            & Measurement\\ \hline
  H/W utilization & This type measures the gap between when resources are 100\% utilized and when optimized by an optimizer \\ \hline
  Cost model      & This type measures the gap between the optimal output of the cost model and the output optimized by an optimizer \\ \hline
  Simulator       & This type measures the gap between an optimal output of the simulator and the output optimized by an optimizer \\ \hline
  \end{tabular}
\end{table}

The first measure is {\em H/W resource utilization} such as CPU.
The strength of this measure is that it is simple and intuitive.
If some optimizers were able to reduce the number or/and the duration of blocking CPU, for instance,
the CPU utilization would increase up to 100\%, defined as the best-case.
H/W resource utilization is a good measure when one wants to optimize I/O-related algorithms or jobs,
because I/O request is a major cause of CPU blocking.
Shafer \ea~\cite{Shafer:2010:HDF} and Li \ea~\cite{Li:2011:PSO} used this type as a measure for optimization.
However, the main drawback is the difficulty of capturing CPU-related overheads.
For example, using only H/W resource utilization data,
it is non-trivial to capture the context-switching overhead, a reducible CPU-related overhead.

As a second type of measurement, Starfish~\cite{Shivanth:2011:PWC} is a well-known optimizer which uses a {\em cost model} of \haa.
The strength of Starfish lies in its flexible cost model, defined as a function with parameters,
representing diverse factors that affect the processing time of a job.
As such, its cost model is viewed as a measure for ``optimization."
If one is able to find a set of parameters that make the output of function (for cost model) minimum,
then one views the output of the cost model as optimal.
To build the cost model, \map\ task and \reduce\ task are divided into
the {\it Reading}, {\it Map Processing}, {\it Spilling}, and {\it Merging sub-phases}
and {\it Shuffling}, {\it Sorting}, {\it Reduce Processing}, and {\it Writing}, respectively.
The cost model of the Starfish consists of the sum of the processing times of these sub-phases.
The profiler of Starfish collects the processing times of these sub-phases
and uses them as raw data for the cost model.
Nevertheless, Starfish has several weak points.
The first is its over-estimated profile processing time of sub-phases in a task.
The second is its sometimes misleading analysis about the best-case scenario which makes the cost model optimal (when it is not in reality).
One of the reasons for these problems is their assumption
that all sub-phases are running on a single thread,
can be too limiting in some scenarios.

Finally, Wang~\cite{Wang:2012:EMS} proposed a {\em simulation-based} approach.
It is similar to Starfish in that it also builds a model for evaluating \ha performance.
As with Starfish, the optimization in Wang's paper means finding a set of parameters
that minimize the output value, expected processing time of a job, of its simulator.
However, there are also differences between the two approaches.
Starfish abstracts a few variables from many factors
that affect the processing time of a job, especially H/W related factors,
in order to make the cost function simple.
In contrast, Wang designed a comprehensive simulator
that consists of sub-systems emulating sub-modules of the \haa.
This approach, while more comprehensive, lacks the
capturing of the cost that occurs between sub-systems.

Overall, these model-based approaches do not suggest independent optimization measures.
In order to evaluate the accuracy of their cost model, therefore,
they have to compare the best case of the cost model and other cases, \eg, the rule of thumbs case.
Of course, after verifying the accuracy of the model theoretically and empirically,
one is able to build a simple measure using the minimum runtime from the best-case.
Considering the complexity of the \ha implementation, though, this is not easy.

\section{Conclusions}

We investigated the question of ``How good is a given \ha optimization?"
by estimating the limit of \ha performance accurately.
Our proposal shows how much more room exists for further optimization, very useful information for \ha practitioners.
To achieve this goal, we analyze the impact of H/W factor toward \ha performance and the cause of overhead.
Then, based on the findings, we propose a novel measure, $vet_{job}$,
to estimate the degree of optimization of a \ha job such that a job with $vet_{job}=1$ is viewed as the perfectly optimized job.
Further, real experiments performed on diverse cluster settings and \ha configurations,
we demonstrate the effectiveness of our proposed measure, $vet_{job}$, with respect to its accuracy,
and discuss how to use our proposal to complement existing solutions such as Starfish.





\bibliographystyle{elsarticle-num}
\bibliography{vet}







\end{document}